\documentstyle[epsfig,aps,12pt]{revtex} 
\begin{document} 
%
\title{Discrete symmetry and quark, lepton and vector
gauge boson mass expressions.}
\author{G.R. Filewood.}
\address{School of Physics,
University of Melbourne,
Parkville, Victoria 3052 Australia.}
\maketitle
\begin{abstract}
Some interesting mass expressions for
all the quarks, leptons and gauge bosons
are presented based on simple discrete symmetry.
 Precise expressions  for the nucleon masses
are presented as an example of the calculation 
techniques employed. The structure of
the Higgs sector is  explored.

\em{Pacs Nos. 12.15.Ff,14.20Dh,
14.60.-z,14.65.-q
,14.70.-e,14.80.Bn.
Paper also available as Revtex document.}
\end{abstract}

$\;$
\section{introduction}

In the standard model the masses of the
fermions are generated by interaction with the
Higgs field;
\begin{equation}
{\cal{L}}_{\mbox{mass}}=g
{\bar{\psi}}_{L}\psi_{R}\phi
\label{mass}
\end{equation}

\noindent
where g is a (dimensionless)
coupling constant and $\phi$
is the scalar Higgs field of
dimension $l^{-1}$. The Dirac mass so
defined is in fact specified by the parameter
g and it is fair to say the expression (\ref{mass})
tells us nothing at all about the masses of the
fermions; the detail is hidden in the arbitrary
parameter g and the unspecified mass of
$\phi$. One might reasonably ask whether
the Lagrangian field theory approach is even  
 {\it{in  principle}} capable
of explaining the masses of the fermions. 
Certainly the conventional perturbative approach
to (electro-weak) field theory would seem to be inadequate
for the purposes of defining particle masses 
given the requirement for mass renormalisation
in such an approach and even non-perturbative
approaches to the strong interaction require
the insertion `by-hand' of the current quark
masses. These failings certainly suggest that
an alternative approach to determining the
origin of the spectrum of particle masses 
might be required; provided always that 
 consistency with the  standard model
is maintained as the latter is certainly 
empirically valid over a wide range of
experimental parameters.

The data on particle
masses is particularly impressive on two counts;
firstly that so much precision data is available
now and secondly that the standard model fails so
miserably to account for  it! It is almost accepted
wisdom in contemporary high energy physics that,
possibly with the exception of neutrino physics,
experiment lags behind theory; yet this is not true
of the data that
 exists on particle masses! It therefore behoves us
to look closely at the data for clues that might alert us
to underlying symmetry not apparent from the standard
model. 

This paper presents such a approach;
it does so by looking at possible 
discrete symmetries which might underpin the
observed spectrum of particle masses and
relating these back to the known 
continuous symmetries of the standard
model. The link between the discrete and
continuous symmetry in the model presented 
is provided by  local gauge invariance.
Particle rest masses then arise as a relic feature
of an otherwise unobservable discrete symmetry
underpinning quark and lepton structure.

Thus the model presented here is essentially
semi-empirical but
 there is a long history
of discovery in physics based on finding
patterns in physical data. Some of the
most famous would include the discovery of
the periodic table of elements and the
Balmer series {\cite{balm}} although one
could argue that all basic science evolves ultimately
from a study of the  patterns observed in nature.

In the framework of the symmetry used to derive
fermion masses all interactions are treated
 in an entirely non-perturbative 
manner. Radiative corrections to particle masses
are calculated  
using a reasonable non-perturbative ansatz.
This gives quite  precise mass predictions  which
can be  compared with experiment;
including the masses of stable hadrons.

The paper is organised as follows;

Sections II-V Introduce the basic geometric ideas 
employed throughout. These sections are long and
quite `wordy' but the concepts are unfamiliar and
a little abstract so it was thought, at the risk of
losing the readers attention,  probably worthwhile
trying to get the ideas across with as much explanation 
as possible.

Sections VI-X start on some more  concrete analysis and
puts the ideas presented in the earlier sections on a
 footing capable of leading to systematic
mass calculations concentrating on first generation
hadrons. The material is quite different to
 a standard QCD approach although the reader will note
some overall symmetry features in common with QCD.
  As mentioned, there is reason
to believe that a standard QFT approach cannot (even
in principle) determine mass ratios from first principles
(i.e. without arbitrary parameters).

In section XI we look at   the Higgs
sector of the theory 
from the point of view of geometry
leading to the masses of second and
third generation objects and massive vector gauge bosons.
This section is quite heuristic as only vague hints of
the origin of the pattern of Higgs couplings can be gleaned 
from the values of the couplings themselves -
which have been determined by comparison with experiment -
although the 
symmetry of the system leads to many constraints on the values
the couplings can take; in particular they can only take 
certain fixed values which then allows comparison with
precision measurements.
Lastly we look at the issue of neutrino masses and 
generalisations of postulate 1 to include more complex models
of the space-time continuum.

\section{background ideas}

Experiment has unequivocally established the reality
of quarks, behaving as point-like objects,
 as the constituents of hadrons and
also shows no indication that leptons have any
sub-structure. Quarks and leptons come in three
generations or families;

\begin{center}
\begin{tabular}{|c||c|c||c|c||c|c|}\hline

quarks;&
top&bottom&charm&strange&up&down\\\hline
leptons;&
$\nu_{\tau}$&$\tau$&$\nu_{\mu}$&$\mu$&
$\nu_{e}$&e\\\hline
\end{tabular}
\end{center}
\noindent
and the masses generally decrease from left to right;
with the exception of the neutrinos which
have much smaller, or zero, mass in comparison
with their partner charged lepton. The other
pattern evident is that, for each family, the
quark masses are greater than the corresponding
charged lepton. Thus the bottom quark, for example,
has a mass in the vicinity of 5GeV whilst the tau
lepton has a mass in the vicinity of 1.7Gev.
In the standard model the masses of the neutrinos are
zero; these being left-handed particles only and
not possessing a Dirac mass. Current data suggests that
neutrino masses are non-zero and we will look at this
issue towards the end of the paper. In what follows
in the interim we will treat the neutrinos as massless.

The standard model theory of interactions is based on
the principle of local gauge invariance. In the case
of electromagnetism this is simply a U(1) phase. 
A  renormalisable Lagrangian is generally taken to
 contain at most the field and
its first derivative only. The 
derivative of the spinor field is not invariant
under {\it{local}} phase transformations;

\begin{equation}
\partial_{\mu}(e^{i\theta_{(x)}}\psi)\,
=\,\partial_{\mu}{\theta}_{(x)}\,
e^{i\theta_{(x)}}\psi
+
e^{i\theta_{(x)}}\partial_{\mu}\psi\;\;
{\neq}\;\;e^{i\theta_{(x)}}\partial_{\mu}\psi
\label{local}
\end{equation}

\noindent
but under a covariant derivative 
containing the electromagnetic field the
fermion field transforms as follows;
\begin{eqnarray}
{\cal{D}}_{\mu}(e^{i\theta_{(x)}}\psi)
\,
&=&\,i\partial_{\mu}{\theta_{(x)}}\,e^{i\theta_{(x)}}\psi
+
e^{i\theta_{(x)}}\partial_{\mu}\psi
+
ieA_{\mu}e^{i\theta_{(x)}}\psi\label{tree}\\\nonumber
\cal{D}_{\mu}&\equiv&\partial_{\mu}+ieA_{\mu}
\end{eqnarray}
\noindent
Now, under the (commuting) U(1) symmetry we 
let the gauge field transform as follows;
\[
A_{\mu}\;{\rightarrow}\;A_\mu^{'}\;{\equiv}\;
A_\mu -{1\over{e}}\partial_{\mu}\theta_{(x)}
\]
\noindent
Finally, substituting this definition in eq.(\ref{tree})
gives;
\begin{equation}
{\cal{D}}_{\mu}(e^{i\theta_{(x)}}\psi)
\,
=\,
e^{i\theta_{(x)}}\partial_{\mu}\psi
+
ieA_{\mu}e^{i\theta_{(x)}}\psi
=
e^{i\theta_{(x)}}({\cal{D}}\psi)
\label{covar}
\end{equation}

In the case of the weak interactions the 
local gauge invariance is more complex but the basic
idea is the same; the spinor field is represented
by a left-handed doublet and right handed singlet.
The corresponding phase factor for the doublet is
an S.U.(2) symmetry and  local gauge 
invariance occurs  under a covariant derivative involving 
a trio of gauge fields derived from the adjoint representation
of S.U.(2); rather than the single field that arises
from the U(1) phase factor in electro-magnetism.
 Thus the right handed singlets do not `feel'
the force mediated by the weak vector gauge bosons. 

Likewise, in the case of quantum chromodynamics,
local gauge invariance under S.U.(3) colour
leads to a theory of the strong interaction with
8 gauge bosons reflecting the adjoint representation.

There is now ample evidence to assert that 
the $S.U.(3){\times}S.U.(2)_L{\times}U(1)_Y$ standard model
is a (not necessarily the) correct description of fundamental
structure yet nowhere do we find, apart from
the subjective criterion of mathematical elegance,
any reason why nature has chosen local gauge invariance
as the cornerstone of its foundation. In the following 
sections we explore
a geometric model capable of accounting for nature's 
predilection for local gauge invariance.

Let us begin by considering the structure
of the continuum. In physics we routinely assume that
space-time is a real continuum. By this we mean
that any interval in space (time) may be represented
by the real numbers as the ratio of the
size of the interval in question  with
some standard unit interval (such as a metre).

The structure of the continuum is quite  
interesting mathematically. In terms of set
theory Georg Cantor showed that the continuum may
be analysed in terms of cardinality, or order,
of sets. This idea will be crucial for our study
of particle masses. We may analyse the continuum in
terms of a hierarchy of cardinality. The first tier
is simply that of a finite set; for example the
set of two integers \{0,1\} (which is also 
the basis of a number field of
 finite cardinality 
called the Galois field). The next tier is
of transfinite (i.e. {\it {completed}} infinite) cardinality
and it is the field of rational numbers (which of course has
the same cardinality as the counting numbers). This order of
cardinality is represented by the symbol $\aleph_0$. Cantor 
proved that $\aleph_0$ is not the cardinality of the real
numbers and is in fact a smaller quantity 
than c the cardinality of the real continuum; in spite of the
fact that $\aleph_0$  represents a (literal) completed infinity.

A subtlety arises with respect to the definition of the
third tier of cardinality due to the fact that it
is not possible to prove whether there exists an order
of cardinality $\aleph_1$ such that 
\[
\aleph_0<\aleph_1<c\]
where $c$ is the cardinality of the real numbers.
One is at liberty to assume the continuum hypothesis,
which asserts that no such number $\aleph_1$ exists,
in which case we define the real number continuum in 
terms of three tiers of cardinality
(i.e. finite, transfinite order $\aleph_0$
and transfinite order c). Alternatively one 
may consider models of space-time based on a continuum
built up from more than three orders of cardinality
in which case one considers that the number 
(or numbers) $\aleph_1$
exists. We will consider both possibilities in this paper.
Initially, however, we will work with the continuum hypothesis
as a postulate;

\underline{Postulate 1};
That there exists an isomorphism  between the
structure of the real number continuum and the
structure of physical space-time which may be
represented in terms of three tiers of cardinality
(the continuum hypothesis).

\underline{Hypothesis 1};
That the fundamental structure of matter and 
energy in the universe is related to a deconstruction
of the the continuum of space-time.

Hypothesis 1 implies that mass/energy and space-time
are made out of the same basic `stuff'.

\section{Affine set geometry}

We define  {\it{affine-set geometry}} as one in which
distances and angles are not defined but only the
shape is manifest. We define an  {\it{affine-set}}  as
a collection of points interconnected by lines.
A point is always a termination of a line. Multiple lines
may terminate at a single point. No points exist on
any line  except at the terminations of the line.
 There are thus always two and only two points associated
with any single line.

Affine-set geometries can be used to model local
gauge invariance but their significance in this paper
is much broader.

Consider for a moment the common geometric idea of a
two-dimensional sub-space of our three-dimensional
real space (time); that is, a two-dimensional plane.
Conventionally we would consider that the physical properties
of the continuum (of 3-space) carry over
 to any two-dimensional
sub-space but in fact we don't really know if this is
true (assuming always that the 3-space of space-time
is in fact a real continuum!). The reason we don't know
is simply because such a space  is not empirically 
accessible. Let us consider this statement.
 For a two-dimensional plane to be truely two-dimensional
its' `thickness' must literally be zero! We can only 
ever assess the properties of any space indirectly; that
is, by observing the behaviour of a test particle
 in that space where
the continuum of an objects  motion, momentum etc.
 implies the
existence of the continuum property of the
underlying space-time. In the case of a true
 two-dimensional space (+time)
it would require an infinite amount of
energy to confine the wave-function of any massive 
 object to the
 space so such observations can never be made.
(For example, to compress length contraction to
zero we would need v = c;
\[
{l^{'}}={l\;{\sqrt{(1-{{v^2}\over {c^2}})}}}
\]
or that the velocity of the massive object reaches the speed of 
light which is impossible).
 Thus we do
not know whether such a two-dimensional 
space would have the continuum 
property of the real numbers or not! we simply cannot confine 
test particle to the space.

The situation with a truely one-dimensional sub-space
is even worse for, in addition to requiring an infinite
amount of energy to confine the wave-function of a massive  
object to a space of zero transverse dimension, the 
line would need to be of infinite length to accommodate 
it! (It is important here to realise that a quantum object
with non-zero rest mass 
cannot be squeezed into a point of zero dimension!).

These statements, however, are not necessarily true
for massless objects such as photons and gravitons
(although, once again, if they have non-zero energy
they cannot be squeezed into a space of zero dimension;
i.e. they cannot be squeezed to a true geometric `point').

One of the central tenets of this work is that any one or
two-dimensional sub-space of our physical 3-space is not
in fact a real continuum and that the continuum property
of space-time only exists at the level of three space
dimensions. Put another way, such sub-spaces do not
exist! But there is a caveat to this assertion; such
sub-spaces may exist as the underpinning of the structure
of fundamental particles themselves; not so much
 as a kind of `internal'
space as a partial `deconstruction' of continuum space.
The role that local gauge invariance will play is to
adjust the cardinality of these one and two-dimensional
subspaces, that are postulated to underpin the structure
of fundamental matter, to that of the continuum. The 
non-observability of any local phase then becomes equivalent
to the unobservable nature of the subspaces `building' the
particle structure.

This is the use to which we wish to put affine geometry.
Consider the idea of  embedding  affine-set geometry
into a continuum space; that is, we want to see 
what the geometrical consequences are of trying to
wed affine-set geometry and continuous geometry.

 For example,
consider the following two (affine-set) triangles
embedded on a continuum circle;

\vspace{0.6cm}

{\centerline
{\epsfig{file=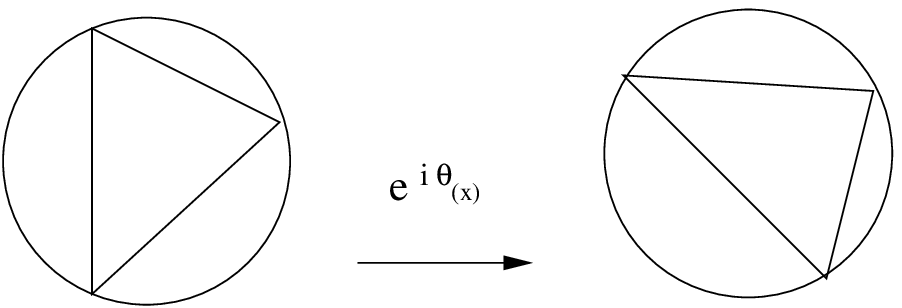,width=10cm}}}

\vspace{0.6cm}

The two triangles represented by these diagrams
are, in affine-set geometry, equivalent. This
is because the side lengths (the three 
`lines' of the triangle) and the angles 
subtended by the lines where they form points
 are not measurable
 by definition  for an affine-set. They do not represent
a physical observable.
  The transformation
which takes one into the other is a local
U(1) gauge transformation on the embedding 
(continuum) circle
however
 as the vertices shift
around the margin of the circle by different amounts
(by contrast a {\it{global}} transformation would 
preserve triangle angles and side lengths). Thus 
 an affine-set triangle embedded in a circle
defines a U(1) local gauge symmetry in the sense that
such a triangle is {\it{invariant}} under both global
and local U(1) gauge transformations
only if the geometry defining the triangle
is an affine-set. In this sense we could in fact use
local U(1) gauge symmetry as an alternative  definition of
affine-set geometry (we don't have to use a triangle of
course - we could use any kind of affine-set geometry).
For example, from the set (infinite but countable) {$\aleph_0$}
of points on the circle we choose an infinite set
of sets of points each with three elements $\aleph_3^i$
where  the superscript i indicates the (countable)
ith set of three points (the cardinality of this set is
of course also $\aleph_0$). Now, assume that each ith set
of points is unique; that is
\[\aleph_3^i\;{\cap}\;\aleph_3^j=0;\;\;\;{\forall}\;i,j\]
This defines an infinite, but countable,
 set of geometric triangles 
inscribed within the circle. Now the transformation
which converts any given triangle into any other
is always a local U(1) transformation (global U(1)
transformations are a subset of the set of 
local transformations). For the total geometry to
be invariant with respect to such U(1) transformations
requires that the geometry be affine-set and can be used as a 
formal definition.

The union of the sets $\aleph_3^i$ defines the set of points
in the bounding circle\footnote{treating
 the circle not as a true continuum of
points since  cardinality 
$c\neq{\aleph}_0$}.
Let $C_3^i$ be a given affine-set triangle. Define the
transformation $C_3^i{\rightarrow}C_3^j$ as $C_3^{ij}$
then;

\begin{equation}
\Sigma_{i,j}(C_3^{ij})=e^{i\theta_{(x)}}
\label{tran}
\end{equation}
\noindent
gives us a compact way of expressing the equivalence 
of a local gauge transformation and an affine-set 
transformation on the circle.

We could add a requirement of `smoothness' to
the affine-set transformations to eliminate singularities
produced by vertices `crossing-over' each other
in the transformation process (as two vertices
pass each other they define a line not a triangle).
Does equation (\ref{tran}) remain valid with the
elimination of such singularities? Indeed, it would seem
that, if $\theta_{(x)}$ is required to be  smoothly
differentiable then such transformations must in 
fact be eliminated and it is a hidden assumption
of eq.(\ref{covar}) that the local phase transformation
$e^{i\theta_{(x)}}$,
like the global one, is smoothly differentiable.
 It is not at all clear
if local gauge invariance is maintained if the transformation
is not a smoothly differentiable manifold!

A way of thinking of this is in terms of an order of the
points in the circle. Treating the points in the circle
as an ordered set (although infinite) we know that we can
still squeeze or stretch any segment of the circle
to an arbitrary degree
without changing the order of the points and thus
preserving the differentiability of the manifold. This follows
from the work of Georg Cantor who proved 
that there is a one-to-one mapping of the points on
any two finite length line segments (and indeed onto
an infinite length line segment from a finite one!).
That a global rotation is a one-to-one order-preserving
mapping is
obvious. It is not so obvious that this is the case 
for a local U(1) transformation however since singularities
can be introduced by such a transformation as mentioned.

Affine-set geometries, because the `length' of sides
of the geometry can only ever be defined as `unity',
can be represented by
   discrete group symmetry. Discrete groups can
be `spinorial' or `bosonic' in the sense that
they contain (res. do not contain) improper rotations.
An example of an improper  rotation in two dimensional
space would 
be;
\vspace{1cm}

{\centerline
{\epsfig{file=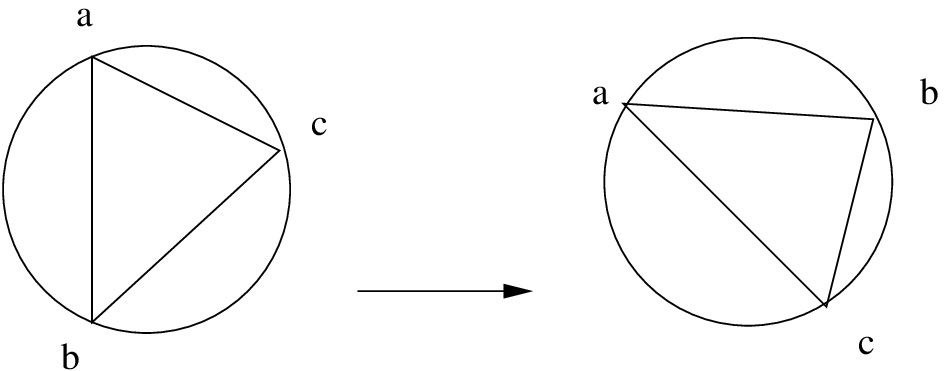,width=10cm}}}

\vspace{1cm}

\noindent
where the vertices `c' and `b' have interchanged
as well as shifted around the circle with a local U(1)
transformation. In this case it is not clear that
the transformation could
 be represented in the form $e^{i{{\theta}_{(x)}}}$ with
continuously differentiable $\theta$ as the set of 
transformations is not one-to-one and onto and 
not well defined (we would not know, for example,
 how many points 
in the circle are crossing over other points
 during such a transformations or indeed if the
 continuum remained  well defined if a (?completed)
 infinite number of cross-overs occurred). This is obviously
 not the case however if the order of the points on the
triangle (a,b,c clockwise or anti-clockwise) is preserved
since we may define the transformation in that case as
purely a stretching or compression of the continuum which
preserves the order of all points in the circle. Thus, if we 
take only the $C_3$ generator of the triangle as being
manifest, and allow it to become a {\it{continuous}}
rotation, we could use the affine-set triangle as a model
of a quantum object of integral spin satisfying a local
U(1) symmetry invariance.

\section{deconstructing space-time to make matter}

We will do this geometrically by making geometric
models of orders of cardinality which we will then use 
simultaneously as  models of
 the space-time continuum and as models of particle structure.

We start with an object to represent the Galois Field
of finite cardinality. This is the discontinuous set
of two digits \{0,1\} which we may represent as 
a discrete one-dimensional interval. Its' only definable
`length' is 1; in units of itself (i.e.; it is unmeasurable)
and the interval possesses an $S_2$ discrete symmetry because
the choice of labelling its' ends with the digits
0 and 1 is an arbitrary choice. Only two geometric
points are associated with the interval; those at
the terminations of the interval. There are no
`points' along the line itself; this is an affine-set
of finite cardinality (cardinality 2) 
and the `line' is absolutely
discontinuous. If we embedded such an
object in conventional space-time it would have a true
`gauge' invariance in the more literal meaning of the word;
that is, it would have the property of {\it{scale}}
invariance. (It is also interesting to note that the
$S_2$ generator, expressed as a continuous geometric
symmetry, would render the object as  spin 2).

To make a geometric model of 
the next highest order of cardinality, that of
 $\aleph_0$,
we must make some further assumptions. Intuitively we
might look at the continuum as rather like a complex structure
built up from simpler `building blocks'; the simpler building
blocks should underpin the more complex structures. We are
working under the assumption that these
`building blocks' are the tiers of cardinality
implicit in the structure of the continuum. Thus
we want to use the Galois geometry to `generate' a space
with cardinality $\aleph_0$. If we regard the Galois geometry
as a `separation' of two points the obvious geometric option is
to iterate this process and separate two Galois intervals
to create a two-dimensional space bounded by Galois intervals.
Such a process can define (at least)
two types of objects; triangles and
squares each object bounded by discontinuous Galois intervals.
 We now assume that the two-dimensional space
contained within the bounding intervals has the cardinality
$\aleph_0$.

Finally to create a model of the continuum we iterate the
process again. This allows us to generate cubes, tetrahedrons
etc. The contained space is assumed to be a 
geometric representation of the real number continuum whilst the
boundaries of the geometries remain discontinuous. Of course, no
{\it{time}} dimension is defined by this procedure; only a three
dimensional continuous but finite space is defined
 contained `within' the boundary of the defining geometry.
Let us call these three-dimensional geometries `Cantor'
geometries.

\section{overview of phenomenological interpretation of the 
geometry linked to local gauge invariance}

{\underline{Hypothesis 2}}
That the Galois interval is the geometry of the graviton.

{\underline{Hypothesis 3}}
That the Cantor triangle is the geometry of the photon.

{\underline{Hypothesis 4}}
That the Cantor cube is the geometry of a quark.

{\underline{Hypothesis 5}}
That the Cantor tetrahedron is the geometry of a lepton.

{\underline{Hypothesis 6}}
The boundary of an object constitutes its' associated
gauge fields.

{\underline{Hypothesis 7}}
Any geometry defining a real-continuum of space
has a mass specified by the symmetry of the object.

{\underline{Hypothesis 8}}
Objects which do not define an intrinsic  continuum 
 to be embedded in space-time must 
elevate their cardinality to that of the continuum.

The  are compelling  empirical reasons for
making these identifications as we shall see but
at this stage our chief motivation is to account for 
nature's obsession with local gauge invariance.

Hypothesis 7 is the basis of rest  mass calculations. We will
form  matrix representations of the symmetry groups
implied by the affine-set; which in general is a permutation
symmetry of a labelling of the points of the set. We will
then sum over all matrix elements needed to describe the
geometry, and, in the case of the strong interaction,
any matrix elements describing the geometry of the
 interaction, and form ratios of particle masses by
taking ratios of sums of matrix elements.

In summary we have assumed that the continuum of space-time
has a complex structure in terms of tiers of cardinality and
that material objects in the universe are a reflection
of that structure.
{\it{Time}}
must be assumed to exist outside the boundaries of the 
geometries in question and may be added as an additional
hypothesis if desired (which is reasonable given that
the existence of time is always assumed in any physical
theory!). We could alternatively define time in terms 
of a relationship between different geometries embedded
in a larger three-dimensional space. The time dimension is of course 
assumed to satisfy the continuum criteria. However,
the cardinality of four-dimensional 
{\it{space-time}} is not assumed to be different
from that of the three-dimensional space as defined above
(in section XI we will look at alternative models of the 
continuum where this latter assumption does not hold).

More particularly the sequence of geometric construction 
must be viewed as the generation of space-time itself. Thus we
do not view the Galois interval geometry as an object
embedded in space-time but rather, in combination with
the two and three-dimensional geometries, as the generator of 
space-time itself. The time-propagating Galois interval will
`sweep-out' a sheet of two-dimensional space. Curvature of this
`swept-out'
space would then be sufficient to elevate the cardinality of
the generated space to that of the continuum under the continuum
hypothesis (viz. postulate 1). However, the one-dimensional
Galois interval, which we have identified with the graviton,
does not have a continuum structure itself; nor does the
`swept-out' sheet of two-dimensional space produced by its'
propagation. Only the final three-dimensional generated space
has the continuum property.

These hypotheses contain a lot of interesting 
implications. Firstly we note that, since all
geometries have Galois intervals in their boundaries,
all geometries, including the photon and the graviton
itself (by hypothesis 6),
 have gravitons as an associated  field
and `feel', and are the source of, gravitation.

An object such as a Cantor 
tetrahedron must be electro-magnetically
charged since it has a photon (gauge field) in its boundary
(i.e. cantor triangles in its' boundary). Let us look at this
geometry from the perspective of local gauge invariance.
The Galois intervals defining the Cantor tetrahedron
are by definition unmeasurable by any observer.
 The terminations of the
intervals can be anywhere in space leading to a
field theory rather than a theory of a point particle.
However, there is an inbuilt paradox in the theory because
the time dimension cannot be defined within the interior of
the tetrahedron; which leads to the object behaving like a 
geometric point in scattering experiments.
 In other words, the Cantor tetrahedron
is a three-dimensional space object embedded in a four-dimensional
space-time. No observer probe can penetrate the 
timeless interior space 
of the Cantor tetrahedron (all observations occur in time!)
which will always appear smaller than the wavelength of any
probe. 

Of particular interest is the boundary between the
timeless three-dimensional space 
in the interior of the Cantor tetrahedron 
and the four-dimensional space-time
outside the boundary of the tetrahedron.
In the case of the Cantor tetrahedron this boundary is composed
of Cantor triangles and Galois intervals defining these triangles.
These are the postulated geometries of the photon and graviton
respectively. Because of the unmeasurable nature of the 
`lengths' of these intervals their terminations can
be anywhere in space and interaction with them 
will be probabilistic; they will manifest as 
fields in space-time (by hypothesis 
 respectively the electro-magnetic
and gravitational fields of the Cantor tetrahedron). By
virtue of the fact that these bounding geometries in
turn define the Cantor tetrahedron so too the Cantor 
tetrahedron will have  the schizophrenic identity of both
a non-local field and a geometric point under observation.
This is a perfect candidate for a quantum field.

One obtains the concept of local gauge invariance
under the more demanding hypothesis 8. Hypothesis 8
is a very natural requirement. In essence it demands
that only the continuum is an observable. The space
contained within the boundary of the Cantor triangle
is NOT a continuum by definition
(it has cardinality $\aleph_0$ {\it{not}} c) and thus is not an
observable space. Its lack of observability is satisfied
physically
if it  propagates at the speed of light
so that the contained space of the Cantor triangle
cannot be placed in the same inertial frame as any
observer. Thus the Cantor triangle will always generate
a three-dimensional space with respect to any observer
by virtue of its relative motion
sufficient to define a continuum space as required
for physical existence under hypothesis 8. The astute
reader will have realised that this requirement is 
equivalent to the principle of special relativity
if one identifies the Cantor triangle with the photon.

\vspace{1cm}
{\centerline
{\epsfig{file=fignew.eps,width=10cm}}}
\vspace{1cm}

A flat two-dimensional square, the model for an
electron  neutrino in this schema,
has no triangle in it's boundary and is thus not
electro-magnetically charged. (It nevertheless has a piece
of phase information in its' boundary which induces a weak
interaction gauge field as we will later see - we will
also look at the issue of different generations of quarks and
leptons
later).

The principle of special relativity emerges quite naturally
from this schema viz. hypotheses  3 and 8. If a triangle is
a photon then, to exist in space-time, postulate 8 tells us
that it must propagate {\it{in time}} to `sweep-out' a three
dimensional space to define a real continuum. To be defined
geometrically in this schema it must sweep out a bounded 
space - which means its' velocity must be finite relative to
cubes and tetrahedrons but never zero. If one adds relative
motion between different cubes and or different tetrahedrons
in a larger space-time embedding then it follows that 
the photon's velocity relative to any given cube or tetrahedron
is always a finite constant. Of course the same must apply
to the Galois interval - the model of the graviton in this
schema - it must propagate in time  and sweep out a volume
to exist; but it does not have to propagate at the
same velocity as the photon! It may of course do so but
there is no prima facia reason in this geometry why a graviton  must
propagate at the speed of light.

A cube in this schema satisfies hypothesis 7 and so should
represent an object with a non-zero rest mass. By postulate
3 it should not carry an EM charge but this is under the
assumption that the six 4-vertex faces of the cube are
two dimensional squares rather that three-dimensional
tetrahedrons; allowing for the latter includes
 the possibility that the cube
can carry unit or fractional charge in addition to being
uncharged. In the scheme that follows all permutations 
allowed by the symmetry appear to manifest physically.

\section{spins and other things}

{\underline{Hypothesis 9}}
Discrete symmetry groups  specified by
any given geometry manifest as an
analogous  continuous symmetry
when the time dimension is added to the space in 
which the geometry is embedded by virtue of the 
generators of the discrete group being converted
into continuous rotation generators of the corresponding
continuous group.

Thus the $S_2$ symmetry of the Galois interval
becomes a continuous spin 2 geometry with the
addition of a time dimension. (It can only be 
spin 2 because the
symmetry generator of the interval 
- a reflection - is invariant with a rotation
by $\pi$). The triangle becomes a spin 1 object because
the discrete symmetry group of a triangle, $C_{3}$,
has a generator axis orthogonal to the plane of
the triangle whose sign changes with rotation by 
$\pi$ and is unity under a rotation by $2\pi$.
For the case of the tetrahedron we will take as its'
symmetry group, rather than the 24 element 
group $T_d$, the unique
dihedral subgroup of SU(2) which has 24 elements and
two generators. In the continuum limit this becomes 
SU(2). (The generators of the dihedral group are 
only two in number but of the three generators of
SU(2) only two are linearly independent
$\;\;[\tau_i,\tau_j]=-i\epsilon_{ij}^{\;\;\;k}
\tau_k$).

{\underline{Hypothesis 10}}

The mass of a three-dimensional object is found by
extracting the time dimension from the embedding
space and is the cardinality of the corresponding
discrete symmetry group that remains after the 
extraction (modulo any radiative corrections
and adjustments for the generators). 

Thus, in the case of the tetrahedron, extracting the
time dimension means removing the dihedral generators
and the associated group elements of SU(2) apart
from a residual  discrete collection of 22 elements which
represent the {\it{rest}} 
mass of the tetrahedron. 
(Of course removing the time dimension
can only define a rest mass and 
cannot define a dynamical mass because the
latter is meaningless in a timeless space!
What we obtain is a definition of the mass
of an object at rest irrespective of the lifetime
of the object;
this may or may not be a measurable quantity!
Only in the situation where the lifetime of the
object is infinite will it be precisely measurable!).

For a Dirac mass term in standard theory we
have coupling of the left and right handed components 
of the fields;

\begin{equation}
{\cal{L}}_{\mbox{mass}}=
g{\bar{\psi}}^{L}\psi^{R}\phi
\end{equation}

\noindent
and in the limit of zero mass the left and
right handed fields decouple;

\begin{equation}
i\gamma^{\mu}\partial_{\mu}\psi^R=m\psi^L
\end{equation}

The geometric meaning of this in the schema being
presented here can be understood by considering the 
photon geometry (hypothesis 3) even though
it is not a fermion. The triangular boundary
of the geometry is constructed of three gravitons (in
the limit of instantaneous time) but the gravitational
gauge field `smoothes' this out so that the local triangular
phase of the geometry is not an observable. The geometry is
then best represented by a circular two-dimensional disc
with a `fuzzy' margin
and the geometry $C_3$ generator as the continuous U(1) generator
of the circle. The left and right handed orientations of the
generator (with respect to the direction of motion of the
object) are distinct and, because of hypothesis 8, cannot be
combined to form a three-dimensional volume for the photon
at rest with respect to an object such as a tetrahedron or 
cube. This means that the photon must be massless (the photon
may have  finite radiative corrections from is gravitational
gauge field which induce  a non-zero minimum energy
for the photon but  its' {\it{rest-mass}} must nevertheless 
always be zero). Note that in this pictorial model the non-`fuzzy'
part of the `disc' is a quantised space which will always appear
smaller than any probe (in the case of a photon the only
`probe' would be a graviton); it is not a measurable space. The 
`fuzz' is due the associated gravitational field of the 
photon and, in principle, extends transversely out to 
infinity.

Thus the meaning of a term like $g{\bar{\psi}}^L\psi^R\phi$
is that the left and right handed components only combine
in the context of a three-dimensional space geometry
and that a two-dimensional (or one-dimensional) object in
this schema must always have zero rest mass. Similarly,
any spinor geometry whose contained space defines an
intrinsic continuum, such as a cube or a tetrahedron,
will have a Dirac mass.

This is, however,
not the complete picture since we will have to add
radiative corrections to any Dirac mass due to the fact that the
boundary of the electron geometry (by
hypothesis that of the tetrahedron) contains photons and 
gravitons - which must propagate 
by hypothesis 8 to be defined
properly  and in so doing will generate additional
components of particle mass. In QFT these contributions
are infinite due to integration over momentum space
but in the present geometric picture they must be 
strictly finite because the analogue of the `bare'
mass in the geometric picture is the order of the 
underlying discrete group (not including generators);
 which is strictly finite.
(The analogue of an infinite `bare' mass would be the
order of the corresponding continuous group; but we can
eliminate this by throwing away all the elements 
other than those which represent the real underlying 
discrete symmetry). Thus in the case of the 
three-dimensional tetrahedron in O(3,1) space-time
the continuous group is SU(2) of infinite order
but the real mass of the Dirac term is represented
by the 24 discrete sub-group elements modulo the two
generators of this discrete group left as massless
intrinsic spin generators.
 
The reader at this stage may be concerned that
no mass units are specified! How can you have a 
mass of 22? The answer is that all masses are
a relative thing. So, for example, if we define the rest
mass of the proton as `1' all other masses could be 
defined in terms of this unit. Ultimately a unit like
a Kg can be expressed in terms of the summed mass
of a certain number of nucleons and electrons,
modulo binding energies, at a certain temperature.
This, in turn, could be expressed solely in terms of
proton masses (provided we know the binding energies, 
$M_p/M_n$ and  $M_p/M_e$ mass ratios etc). The beauty
of defining mass in terms of proton or
electron rest mass is that, as far as we know, these 
are the only two (?absolutely) stable massive 
colourless fermions. Using the $\Delta{\mbox{E}}
\Delta{\mbox{t}} \geq{\hbar}$ relation means that,
at least in principle, we can measure the mass of a
stable object to any desired level of accuracy.
In the situation at
hand  what we are really interested in is
calculating quantities such as the rest-mass ratio
of the proton to the electron which we shall do 
by performing a similar calculation for the proton that
we have performed for the tetrahedron (here a model of
the electron - we will deal with muons and taus later).

Notice that the approach to particle structure and
groups being employed here is fundamentally different
to that of the standard model; the SM symmetry involves
relations between different types   particles by assuming the 
spectrum of particles is governed by some symmetry group
rather than the intrinsic structure of a given particle being
governed by a symmetry group; the latter
 is what we are doing here.
Notice also that there is no conflict between the model of
a charged lepton (here a tetrahedron) being built up here and
the empirical finding that leptons appear to be point particles
in deep inelastic scattering experiments. The reason is that
time exists only `outside' the boundary of the tetrahedron -
there is no time in the `interior space' of the tetrahedron
and hence no dynamics can occur in the interior space
which is a three-space not a four-dimensional space-time.
 Observations
of the lepton involve interactions with its' boundary only
and the `interior space' has no measurable size.
Experiment can only detect the presence of this
3-space in an on/off sense; the lepton is either
detected or it is not.  Exactly the same
applies to the 2-space defined by the photon; the photon can
only be probed with regard to its' boundary gravitational
field and the interior space is unmeasurable apart from the
sense that it's existence can be equilibrated with detection of
the photon.

By the same token, because the boundaries of the (timeless)
contained spaces are themselves gauge quanta, the boundaries of
the geometries will represent propagating fields and thus 
the whole geometry will represent a field quanta rather than
a static geometric point in space-time. The boundaries of 
geometries such as affine-set tetrahedrons are thus in this sense non-local
in this schema and represent the gauge quanta of the
object. A spinorial tetrahedron in this model represents
an electron. The gauge quanta 
(in the case of the tetrahedron the triangular faces
and individual lines of the boundary of the geometry)
eliminate any observable local
phase associated with  the boundary of the geometry (`smoothing-out'
the triangles and intervals) by propagating in space time and 
by sweeping-out an associated 3-space elevate the cardinality
of the boundary of the geometry to the continuum.
The affine-set tetrahedron  may be idealised
as a three-dimensional timeless fuzz-ball embedded in a four-dimensional
space-time with the surface `fuzz', extending out to
infinity, being the propagating gauge quanta represented
by the surface of the geometry propagating at the
speed of light. The `interior' time-less space 
is impenetrable and will always appear
smaller that the wavelength of any probe which, by the  very
nature of a dynamical probe, is  constrained to lie within the 
boundaries of the time dimension and cannot enter the 3-space
which defines the underlying geometry.
Thus the tetrahedron always appears as a point particle in 
deep inelastic scattering experiments. This is the  
link  to eqn.({\ref{covar}}); the non-observability of 
local phase arises because one and two-dimensional sub-spaces
of continuum 3-space do not represent a real continuum and
are not accessible to observation. Intrinsic particle spin
in the model is then a necessary prerequisite for the existence
of a fundamental quantum object in space-time which leads to
trouble incorporating a scalar Higgs in the theory (needed
as a `free' particle
to preserve unitarity for radiative corrections in the SM
for example).
We will look at this issue again towards the end of the 
paper but note at this stage that, whilst it appears we
cannot express the Higgs as a free-particle in space-time
in this model as so far developed, 
we can give it physical existence in 
one space; the interior timeless 3-space of a geometry
such as an affine cantor tetrahedron. Indeed, we might seek
to identify the 3-space (minus the boundary of the tetrahedron)
{\it{as}} the Higgs field itself. This is in fact the
route we will take. The Higgs field will be `buried outside
time'.

\section{cubes, quarks and fractional charge}

By hypothesis we identify the Cantor cube with a quark. We 
will be concerned here with derivation of the proton and
neutron masses so we will require a representation theory
that encompasses three quarks and the associated gluon 
field. The following components are calculated in sequence;

1; constituent quark mass

2; gluon energy associated with the constituent quarks

3; current quark masses

4; gluon energy associated specifically with the current quarks.

The constituent quark mass represents the energy associated 
with the momentum of the quarks and in the nucleons we will
find that it represents about one-half of the total mass
of the nucleon. The gluon energy, 2 above, represents the
energy associated with the momentum of the gluons and again
will be found to represent about one-half of the total energy
of the nucleon. The current quark mass is a different component
and represents the intrinsic rest-mass energy of the individual
quarks. Component 4 is something of a mystery but appears to be
rather like an instantaneous  strong-force analogue of the
instantaneous Coulomb interaction. It might be viewed as a 
potential energy of separation of the current quarks at the
energy scale of the calculation.

To study the symmetry properties of the Cantor cube geometry
we will need a representation theory which is given below.
The details are somewhat more involved than in the case of
the Cantor tetrahedron but
an essential summary of the ideas is as follows.
Hypothesis 8 requires the propagation 
(at light speed) of all boundaries that
do not define a rest mass 
(which means any geometry that is one or
two-dimensional) and hypothesis 9 means that the
cube boundary will have a complex structure. Hypothesis
8 will mean that a flat square geometry  must be
massless; when confined 
to a colour space (defined below) this  will manifest 
 as a massless spin-1 object  (i.e. as a gluon) 
and the representation theory specifies
that the cubes are fractionally charged but massive.
(The fractional charge results from triangulation  of a fractional
part of the boundary of the cube; geometrically this means
that some of the square surfaces of the cube become a 3-space
rather than a 2-space so that the geometry becomes a tetrahedron
rather than a square - the charge of the cube is then represented
by the fraction of the topology that is triangulated. The group 
theory is such that opposite square faces of the cube always
triangulate in tandem so that the possible fractional charges are
$\pm {1\over3}$ and $\pm {2\over3}$ with the sign dependent on
the orientation of the $C_3$ generator).

  Essentially all
phase information with regard to the boundary of any geometry
is eliminated by the gauge fields; the position and direction
in space of the vertices and edges of the geometry are  not
an observable. In the case of a cube the 
overall `cube' phase information
is its'colour and the three axes of the colour may
be taken as the three (not necessarily orthogonal
in real space but orthogonal in affine-set geometry!)
 lines running through the centre of opposing `square' faces
of the cube. (Observation of the colour of a cube would
be analogous to removing the intrinsic spin of a quark
or freeze-framing the object by removing the time dimension).

We begin by forming a representation for
a spinorial version of a tetrahedral geometry
which will form the basis for a representation theory
of the cubic geometry.
The tetrahedron has special discrete symmetry;
a symmetry related to the 24=4!  elements of a
permutation of its vertices. For a spinorial
version embedded in a 4-dimensional space-time we seek 
the unique sub-group of SU(2) with 24 elements 
which is;

\begin{equation}
\begin{array}{ccccc}
&1&2&3&4\\\\
\alpha:&
\left(
\begin{array}{cc}
1&0\\0&1
\end{array}
\right)&
\left(
\begin{array}{cc}
-1&0\\0&-1
\end{array}
\right)&
\left(
\begin{array}{cc}
0&i\\i&0
\end{array}
\right)&
\left(
\begin{array}{cc}
0&-i\\-i&0
\end{array}
\right)\\\\
\beta:&
\left(
\begin{array}{cc}
i&0\\0&-i
\end{array}
\right)&
\left(
\begin{array}{cc}
-i&0\\0&i
\end{array}
\right)&
\left(
\begin{array}{cc}
0&-1\\1&0
\end{array}
\right)&
\left(
\begin{array}{cc}
0&1\\-1&0
\end{array}
\right)\\\\
\gamma:&
\left(
\begin{array}{cc}
ia&-b\\b&-ia
\end{array}
\right)&
\left(
\begin{array}{cc}
-ia&b\\-b&ia
\end{array}
\right)&
\left(
\begin{array}{cc}
-ia&-b\\b&ia
\end{array}
\right)&
\left(
\begin{array}{cc}
ia&b\\-b&-ia
\end{array}
\right)\\\\
\delta:&
\left(
\begin{array}{cc}
b&ia\\ia&b
\end{array}
\right)&
\left(
\begin{array}{cc}
-b&-ia\\-ia&-b
\end{array}
\right)&
\left(
\begin{array}{cc}
b&-ia\\-ia&b
\end{array}
\right)&
\left(
\begin{array}{cc}
-b&ia\\ia&-b
\end{array}
\right)
\\\\
\epsilon:&
\left(
\begin{array}{cc}
a&ib\\ib&a
\end{array}
\right)&
\left(
\begin{array}{cc}
-a&-ib\\-ib&-a
\end{array}
\right)&
\left(
\begin{array}{cc}
-a&ib\\ib&-a
\end{array}
\right)&
\left(
\begin{array}{cc}
a&-ib\\-ib&a
\end{array}
\right)
\\\\
\phi:&
\left(
\begin{array}{cc}
ib&a\\-a&-ib
\end{array}
\right)&
\left(
\begin{array}{cc}
-ib&-a\\a&ib
\end{array}
\right)&
\left(
\begin{array}{cc}
ib&-a\\a&-ib
\end{array}
\right)&
\left(
\begin{array}{cc}
-ib&a\\-a&ib
\end{array}
\right)
\end{array}
\end{equation}

\noindent
where $a={1\over2}$ and $b=\sqrt{3\over2}$.
The elements of these matrices are built up
from the roots of unity; square roots, fourth 
roots and sixth roots. These elements close to
form a group with 24 elements which we shall
designate $T_r$. The generators of the
group may be taken as the $\gamma_4$ and
$\delta_4$ matrices. Note that;
\begin{equation}
(\gamma_4.\delta_4)^2=(\delta_4.\gamma_4)^2=-I_2
\end{equation}
\noindent
so that, since SU2 is a double covering of SO3
{\it{up to a sign}},
the square of the product of the generators 
represents a rotation by $2\pi$ so that the
$T_r$ group is a kind of discrete spinor group.
It is analogous to the geometric group
Td familiar to physical chemists which is the
group of the tetrahedron in three-dimensions with
24 elements and 2 generators. Td is a point space
group in three dimensions which is spinorial
but is not isomorphic to Tr.

The $T_r$ group will be used as the fundamental
form of a discrete representation for spinor 
particles in the theory.

The {\it{rest mass}} of Tr is defined as ${\cal{R}}.(4!-2)$
where (4!-2) is the number of non-generator elements
in the group and $\cal{R}$ incorporates any radiative
corrections due to the elevation of the cardinality
of the boundary of the geometry to the continuum
by the gauge fields. In any continuous
 field theory analogue the
discrete pair of Tr generators must become continuous
rotations and by hypothesis manifest as the intrinsic spin of the
geometry; this is the reason for defining them as  massless.
The hypothesis of massless group generators
 is testable by comparing resulting mass
calculations with experiment over a variety of particle 
types. The best experimental evidence for the correctness
of this assumption comes from the massive charged leptons which
all have identical Tr group elements 
with identical radiative corrections in their mass expressions.

\section{Embedding $T_r$ in SU(3)}

To calculate mass expressions for cube geometries,
which by hypothesis we equate with quarks, requires
a matrix representation of the symmetry group of the
cube that parallels that developed for the tetrahedron.

We can mirror the root structure of SU(3) by
embedding three copies of  SU(2) 
in SU(3). We can  embed three copies of the $T_r$
group as follows;
\begin{equation}r=\left(
\begin{array}{ccc}
1&0&0\\
0&\lambda_{11}&\lambda_{12}\\
0&\lambda_{21}&\lambda_{22}
\end{array}
\right)
\;;\;g=
\left(
\begin{array}{ccc}
\lambda_{11}&0&\lambda_{12}\\
0&1&0\\
\lambda_{21}&0&\lambda_{22}
\end{array}\right)\;;\;
b=
\left(
\begin{array}{ccc}
\lambda_{11}&\lambda_{12}&0\\
\lambda_{21}&\lambda_{22}&0\\
0&0&1
\end{array}
\right)
\end{equation}

\noindent
and we also form the following `colour-dual'
embeddings;

\begin{equation}\bar{r}=\left(
\begin{array}{ccc}
-1&0&0\\
0&i\lambda_{11}&i\lambda_{12}\\
0&i\lambda_{21}&i\lambda_{22}
\end{array}
\right)
\;;\;\bar{g}=
\left(
\begin{array}{ccc}
i\lambda_{11}&0&i\lambda_{12}\\
0&-1&0\\
i\lambda_{21}&0&i\lambda_{22}
\end{array}\right)\;;\;
\bar{b}=
\left(
\begin{array}{ccc}
i\lambda_{11}&i\lambda_{12}&0\\
i\lambda_{21}&i\lambda_{22}&0\\
0&0&-1
\end{array}
\right)
\end{equation}

\noindent
The three `colour' matrices
$r,g\;\mbox{and}\;b$ plus the `colour-dual'
basis $\bar{r},\bar{g}\;\mbox{and}\;\bar{b}$
together form a discrete version of the 
 group of $SU(3)_c$.
Also notice that each set of colour matrices
forms a group with 24 elements. The `dual' or
`colour-bar' matrices do not form a group
since the product of any two colour-bar
matrices is a colour matrix. However, the
combination of $c_i+\bar{c}_i$ (for colour index
i) forms a group
with 48 elements and two generators located
in the colour-bar set of elements. Let us
call this group $T_c$. It is also a 
`discrete spinor' since the generators
are analogous to the $T_r$ group;
\begin{equation}
(i\gamma_4.i\delta_4)^2=-I_2
\end{equation}
Note that the colour-dual is not the same
as the Hermitian conjugate matrix.

We could consider  $T_c$ group  as a model
for a Skyrmion field of charge 4 i.e.
one with cubic symmetry. We will see 
later how the geometric analogy works
but to obtain a picture think
of the three colours as represented
by the three pairs of opposite faces of
the cube; one pair of faces representing
the $T_c$ group with the four vertices of
each square face representing each tetrahedral
sub-group. To represent the total geometry
we form a `particle vector' which is composed
of three components. First form a matrix
representation of $T_c$ as follows;
\begin{equation}
R=\left(\begin{array}{cc}
r&0\\0&\bar{r}\end{array}
\right),
G=\left(\begin{array}{cc}
g&0\\0&\bar{g}\end{array}
\right),
B=\left(\begin{array}{cc}
b&0\\0&\bar{b}\end{array}
\right)
\end{equation}
\noindent
and then form the three-component particle
vector $\stackrel{\rightarrow}{P}=
(R,G,B)$.
(Of course the order of the components here
is arbitrary).

Now define the following `current-quark' operators;
\begin{equation}
q_r=\left(
\begin{array}{ccc}
-1&0&0\\
0&i&0\\0&0&i\end{array}
\right),
q_g=\left(
\begin{array}{ccc}
i&0&0\\
0&-1&0\\0&0&i\end{array}
\right),
q_b=\left(
\begin{array}{ccc}
i&0&0\\
0&i&0\\0&0&-1\end{array}
\right)
\end{equation}
\noindent

Lastly we require what is referred to in the
text as a `script identity' $\cal{I}$. It carries
a colour index;
\begin{equation}
{\cal{I}}_r=\left(
\begin{array}{ccc}
1&0&0\\
0&-1&0\\
0&0&-1\end{array}\right),
{\cal{I}}_g=\left(
\begin{array}{ccc}
-1&0&0\\
0&1&0\\
0&0&-1\end{array}\right),
{\cal{I}}_b=\left(
\begin{array}{ccc}
-1&0&0\\
0&-1&0\\
0&0&1\end{array}\right)
\end{equation}

The purpose of these constructions will become
apparent as follows. We form operators from the
$q_i$'s and $\cal{I}$'s as follows (the following
examples are red format but the other colours follow
suit);
\begin{equation}
\begin{array}{cc}
U_r&D_r\\\\
\left(
\begin{array}{cc}
{\cal{I}}_r&0\\
0&I_3\end{array}\right)&
\left(
\begin{array}{cc}
q^*_r&0\\
0&q_r\end{array}\right)\\\\
\left(
\begin{array}{cc}
q^*_g&0\\
0&q^*_g\end{array}\right)&
\left(
\begin{array}{cc}
I_3&0\\
0&I_3\end{array}\right)
\\\\
\left(
\begin{array}{cc}
q^*_b&0\\
0&q^*_b\end{array}
\right)
&
\left(
\begin{array}{cc}
I_3&0\\
0&I_3\end{array}
\right)
\end{array}
\end{equation}

\noindent
where each operator has three
components represented above in column format.
The effect of the operators on the colour neutral
particle vector is as follows;
\begin{equation}
D_r(U_g\{U_b
\left(
\begin{array}{c}
R\\G\\B\end{array}
\right)\})=
\left(
\begin{array}{c}
{\bar{R}}\\{\bar{G}}\\{\bar{B}}\end{array}
\right)
=\bar{P}
\end{equation}

Contrast this with the effect of the
conjugate operators;

\begin{equation}
D^*_r(U^*_g\{U^*_b
\left(
\begin{array}{c}
R\\G\\B\end{array}
\right)\})=
\left(
\begin{array}{c}
{{\cal{I}}_r\bar{R}}
\\{{\cal{I}}_g\bar{G}}
\\{{\cal{I}}_b\bar{B}}\end{array}
\right)
={\cal{I}}\bar{P}
\end{equation}
\noindent
which is the definition of unit negative 
electro-magnetic charge
in this scheme; the non-script prefix on
the altered P is the definition of unit positive
electro-magnetic charge.

Two things need mentioning at this stage;
1; the quark operators $q_i$ and ${\cal{I}}_i$
are assumed confined to their respective colour
space i; that is, the $q_r$ operators only operate
on the R matrix etc, and 2; the order of the 
application of the operators does not change the 
result; so long as each operator colour is 
represented only once
(that is $D_gU_rU_b$ gives the same result as
$U_gD_rU_b$ etc).

\section{Glue}

The gluon geometry is represented by a four-vertex
flat (and hence massless) spin-one geometry related
to the boundary of the cube
(the sign of the $C_4$ generator of the
flat square geometry changes sign with a 
rotation by $\pi$ so that the continuous analogue
must be a spin-one generator). There are  six
colour states defined directly from the
cube geometry;  $R,\bar{R},B,\bar{B},G\;\;\mbox{and}
\;\;\bar{G}$. (The remaining generators of SU(3) are 
linear combinations of these).
We can form a representation theory for the 
gluons and we find that their energy equivalence is 
constrained by the symmetry allowing their contribution
to the rest mass of particles to be estimated
by summing over the possible matrix representations
in the same way as was done for the $T_r$ and $T_c$ groups.
There will be  an implicit assumption in such a summation
that $\alpha_s=1$ because we will equate the energy equivalence
of one matrix element of the glue with that of the constituent
quarks. In fact this choice is the only one possible since they
are calculated from the same basis groups and particle vector.

Discretised gluon operators are represented
as follows;
\begin{equation}
b\;\left(\begin{array}{cc}
I_3&0\\
0&q_bq^*_g
\end{array}
\right)
g^*.
\left(
\begin{array}{cc}
g&0\\0&\bar{g}
\end{array}
\right)
=
\left(
\begin{array}{cc}
b&0\\0&\bar{b}
\end{array}
\right)
\end{equation}

\noindent
where * represents the complex conjugate.
Similar upper product 
operators are defined for example;

\begin{equation}
r\;\left(\begin{array}{cc}
q_rq_g^*&0\\
0&I_3
\end{array}
\right)
g^*.
\left(
\begin{array}{cc}
\bar{g}&0\\0&{g}
\end{array}
\right)
=
\left(
\begin{array}{cc}
\bar{r}&0\\0&{r}
\end{array}
\right)
\end{equation}

These are then combined to form 
operators which operate on the particle
vector $\stackrel{\rightarrow}{P}$
for example;
\begin{equation}
\left(
\begin{array}{c}
g\;\left(\begin{array}{cc}
I_3&0\\
0&q_gq^*_r
\end{array}\right)
r^*.\\\\
r\;\left(\begin{array}{cc}
I_3&0\\
0&q_rq^*_g
\end{array}
\right)
g^*.\\\\
\left(\begin{array}{cc}
I_3&0\\0&I_3
\end{array}\right)
\end{array}
\right)
\end{equation}

\noindent
when applied to the particle vector
$\stackrel{\rightarrow}{P}=(R,G,B)$
produces $(G,R,B)$ i.e. this is a R-G
gluon.

\section{radiative corrections}

As was suggested earlier in the text
{\it{bare}} mass is to be interpreted as the transformation
of a geometry over a timeless space or, equivalently,
as a symmetry which represents such  a transformation.
Actually, apart from the absence of the
time dimension,  this is remarkably like the
definition of energy associated with the momentum 
of a moving  object for the motion of an
object is nothing other than its' translation
over a  space in a definable time. Mass and
translational energy thus have a similar 
structure!  For the $T_r$ group,
which represents a discrete spinorial tetrahedron,
there are 24 discrete elements or transformations
over its contained (timeless!) 3-space. 
We have made the assumption
that the two generators of the geometry generate
its intrinsic spin (in the transition to a field
theory) and that these generators are as a consequence
massless. The Dirac mass of the tetrahedral group would
then be given as;
\begin{equation}
massT_r=(4!-2)
\label{barem}
\end{equation}
Eqn.(\ref{barem})
is a finite {\it{bare}} mass. It will be modified
by the electro-magnetic properties of the boundary 
of the geometry. In standard theory component 
(\ref{barem}) 
must be generated by the Higgs field.

That the radiative correction from electro-magnetism
to the mass of an object such as an electron must be
finite and of order $\alpha$ 
was, I believe, pointed out by Dirac and 
has been emphasised by authors such as Sakurai
\cite{sakurai} . The
reason is that, 
for the electro-magnetic contribution ${\delta}m$ 
to contribute {\it{most}} of the mass of the
electron m, the energy of the virtual photons $\Delta$ must exceed
the mass of the entire universe!

\[
{\delta}m={{3{\alpha}m}\over{2\pi}}log\left\{{\Delta\over{m}}
\right\}
\]

 If the E.M. contributions
are of order $\alpha\{{M_{bare}}\}$ and finite then
${M_{bare}}$ must also be finite! Of course in the
 perturbative approach both $M_{bare}$ and the radiative
correction are infinite and the coupling, $\alpha_{em}$
varies with the photon momentum which is integrated
over. The infinities are present even if the momentum
integrations are regularized with a cut-off.

By contrast in the current {\it{non-perturbative}}
approach we have a finite
`bare' Dirac mass. Now, because in the
geometric picture presented the gauge fields are
compressed onto the boundary of the geometry,
we expect on basic symmetry grounds that if the 
Dirac mass can be calculated from the discrete geometry
then so should the radiative corrections! (It might be
argued that this is a rather loose argument though!)
The basic symmetry argument is as follows. There are four
triangles on the surface of the geometry and each has a 
permutation symmetry of 3!=6 elements. Four times six
equals twenty-four so the surface of the geometry,
in terms of photons as triangle affine-sets, has a group
order equal to that of the tetrahedron. We assume however
that any mass-equivalence is reduced from unity to the coupling
strength of the force and that the amount of energy that 
the surface triangles can express is exactly equal to the
bare mass of the tetrahedron because of the equivalence
of the order of the two sets (we assume that the
bare mass is converted to the  photon energy
and that the coupling strength $\alpha_{em}$ is then set
at this level).

 The simplest and most natural 
possible ansat{\"{z}}
is  then given by; 
\begin{equation}
massT_r=(4!-2)(1+\alpha_{(q^2=m^2)})
\label{lepton}
\end{equation}
\noindent
where $\alpha$ is the fine structure constant
defined at the (finite!) bare-mass
energy  m of the particle in 
question. We do not integrate over
photon momentum! We do not use a range
of values of $\alpha_{em}$ but only that 
value set at the energy scale of the underlying
(finite) bare mass. The validity of this ansat{\"{z}}
depends on experimental testing. We will see shortly
that it is accurate to within current experimental limits.
Note that this radiative correction is NOT perturbative.

We next generalise this ansat{\"{z}} to the weak interaction
to accommodate weak radiative corrections to the lepton mass
defined by eqn.(\ref{lepton}) under similar assumptions
 by adding;
 \begin{equation}
massT_r=(4!-2)(1+\alpha_{(q^2=m^2)}+G_f)
\label{leptons}
\end{equation}
\noindent
where $G_f$ is a weak interaction correction
expressed as a dimensionless number in terms of
relative strength in comparison to the E.M. force 
i.e. $G_f\approx10^{-5}$. We will symbolise the
radiative correction as;
\[
{\cal{R}}=(1+\alpha_{em}+G_f)\]

But what is the value of the strong coupling
for the quarks? Comparison with experiment will
show that, for the discrete non-perturbative 
geometric approach, it always takes the value
unity (in dimensionless `relative strength' terms).
In conventional QCD only the asymptotic `free' 
region of phase space is accessible to perturbative
calculations and the low-energy region presents
difficulties (lattice approaches etc). Actually
there is no inconsistency in finding that the
coupling always takes the value unity for affine-set
calculations; if we are not looking at a quark we 
really don't know what it is doing! And if we are 
measuring the rest mass of a proton (ultimately
what we are trying to calculate here) then we definitely
don't know what the quarks are up to for such a 
measurement at rest energy! There are basic
symmetry reasons why the coupling should be unity
in the affine-set regime (it has been called
the `leggo-block' approach) based on the idea
of `stacking' cubes and taking the interaction
energy in terms of the energy equivalence of 
the contacting faces. Because this energy is 
expressed as unaltered matrix elements the coupling is
automatically unity! We will not explore the issue
further but assume that, for systems containing quarks only,
the identity in the factor $\cal{R}$ is an expression
of $\alpha_{strong}$.

Because the model of quarks developed in 
the previous section is based on units of the
$T_r$ group we will make identical
assumptions about the mass equivalence of
respective $T_c$ mass units; that is, we will
equate the mass with the sum  of matrix elements.
 We can write each
$6\times6$ $T_c$ matrix two ways;
\begin{equation}
\left(
\begin{array}{cc}
C_i&0\\
0&\bar{C}_j
\end{array}\right)
\;\;\;\;
{\mbox{or}}
\;\;\;\;
\left(
\begin{array}{cc}
\bar{C}_i&0\\
0&{C_j}
\end{array}\right)
\label{two}
\end{equation}
\noindent
These two possible forms of $T_c$ represent
opposite parity assignments
and the two generators of the $T_c$ group
occur only in the barred matrix.
  As before
we assume that these generators, in the
continuum limit, generate the continuous intrinsic spin
of the geometry and are massless. The two
matrices above will thus  generate, for a single
colour, $2.24.22$ elements of mass as
a result. Now,
summing over three colours gives 
a multiplier of 3 and there are 3! possible
permutations of colour order in the
particle vector. Thus the total number of possible
{\it{different}} matrix elements in an {\it{ordered}}
particle vector (with E.M. and weak radiative corrections) is; 
\begin{equation}
3.3!.2.24.22(1+\alpha_{em}+G_f)
=
6.4!.6.(4!-2)(1+\alpha_{em}+G_f)
\label{order}
\end{equation}
\noindent

Notice that in eq.(\ref{order})
a gauge field correction identical to that
for the $T_r$ group  has been added; the
radiative corrections are based on the
tetrahedral geometry and $T_c$ is just a sum
of such tetrahedral geometries each of which
has its boundary `smoothed' by the appropriate
gauge field with identical corrections to mass 
as for the $T_r$ group. It is reasonable to take
the value $\alpha_{em}$ in eq.(\ref{order})
as the low energy value because the current quark
masses are small for the first generation
(N.B. $T_c$ does {\it{not}} define the current quark
masses - it defines constituent quark energy-momentum).
Notice also that $G_f$ has been coupled to 
both the right and left-hand parts of the particle
vector (expression(\ref{two}) describes opposite parity
states) as was the case for the coupling for the
$T_r$ group. If one wished one could couple a 
value $2G_f$ to half the mass,
as a means of representing the parity violating
nature of the weak interaction,
 but the result would
be the same so we will stick with the current
notation.

 Notice also that in (\ref{order})
the R.H.S. has an interpretation in 
terms of a cubic geometry with
six 4-point geometric surfaces. 
This analogy will be reinforced later.
The double counting implicit here 
(we have the square of the surface order of
the cube; the surface order being 6.4! -
for six four-point subgeometries - 
or 6.(4!-2) depending on whether generators are
included) is 
a reflection of the dual nature of the
$T_c$ which is spinorial and results in
a double counting of the geometry surface.
Mass (\ref{order}) will represent  the 
{\it{constituent}}
mass of a cubic baryon.

To calculate the corresponding 
{\it{gluon}}
constituent mass we notice that
each gluon operator contains the
product $C_i.C_j^{\dagger}$
neither of which contains the $T_c$ 
generator pair. The order of this 
product is thus $24^2$. For a given
three-component particle vector
there are three possible couplings of
colour exchange; first component and
second, second and third and first and
third. This triples the gluon order 
quite independent of the particle vector.
For the particle vector there are three
cyclic permutations which cannot be interconverted
with the exchange of a single gluon
resulting in a further tripling of the 
gluon order. The gluon has two
active colour component in its particle vector
analogue and there is a second doubling of the
order that arises from coupling to the
two different formats in eq.(\ref{two}).
Thus the net gluon matrix order is;
\begin{equation}
{\mbox{gluons}}\;\;\;\;
2.2.3.3.24^2=6.4!.6.4!
\label{glug}
\end{equation}

\noindent
which again has ready interpretation
in terms of the surface of a cube.

We can now sum the $T_r$ equivalent {\it{order}} for the
basic cubic geometry;

\begin{equation}
\mbox{cubic mass}=
\{6.4!.6(4!-2)+6.4!.6.4!\}
.(1+\alpha+G_f)
\label{order3}
\end{equation}

Notice that a radiative correction $\cal{R}$
containing $\alpha_{em}+G_f$  has been 
added to the gluon energy in (\ref{order3})! This seems
completely out of place; an explanation is given later
in the text.

\section{Current Quark Masses}

We first examine the current quark content
of the proton. The current quarks are built
up from the quark operators used previously
to define the particle types. The current quark
masses are the corresponding masses of these
operators plus an extra piece that represents  either (finite)
gluon radiative corrections to the current quark masses
or a potential energy of separation of the current quarks
- we really are not sure quite which!

The glue and quark masses calculated in the previous section
are related to gluon and quark energy-momentum 
due to the dynamical motion of these objects in
the hadron. In this section we
calculate the self-mass of individual quarks;
this is the {\it{current quark mass}}.

We first study the operator component of mass 
of the current quarks. The glue  corrections  are
less well understood. Our aim here is to
extract the $T_r$ equivalent order (i.e. number of $T_r$
equivalent
matrix representations) of the quark operators
in the previous sections.
We can split up the operators into two
pieces; which is convenient for calculations.
Each current quark operator consists of a vector composed
of 3 square $6\times6$ matrices. Each $6\times6$
block is composed of two $3{\times}3$ blocks on
the diagonal. We 
separate out
the upper $3{\times}3$ blocks as one set of
matrices and the lower $3{\times}3$ set of 
blocks as another and represent them as
follows (the representation below is for
the proton as the reader can confirm
by applying the operators to the
particle vector);

\begin{equation}
\mbox{strong component}=
\left(
\begin{array}{ccc}
\cal{I}&q^*&I\\
q^*&\cal{I}&I\\
q^*&q^*&q^*\\
\end{array}
\right),
\;\;\;
\mbox{E.M. component}=
\left(
\begin{array}{ccc}
{I}&q^*&I\\
q^*&{I}&I\\
q^*&q^*&q\\
\end{array}
\right)
\label{cq}
\end{equation}

The rows of these matrices are the three
colours (order red, green and blue in the
notation used in this paper from top down)
and the columns are the type of quark;
the left hand and central column of each
matrix above is an up-type quark and the right
hand column is a down-type quark. The up-type
quarks have their colour defined by the colour
of the single identity they carry. Thus for example
the first column of these matrices represents a 
red up quark. The single q or q* in the down defines its
colour so that the last column above is a blue down
quark.

The strong components couples to the $C$ colour matrix
of the particle vector which does not contain
any generators. The E.M. component couples
to the $\bar{C}$ matrix with two generators.
(This coupling is required to properly define the
charge characteristic of the particle in the given 
representation).
The order of a $C$ matrix is 24 elements. The
order of a $\bar{C}$ matrix is (4!-2) elements. The masses
of the current quark operators are  defined in
terms of the matrices to which they couple.
The exception to the rule applies to the
identities in the E.M. component which are
spinless and of order 4! (i.e. the generators
of $\bar{C}$ are massive under the identity
components of the current quarks - the I
or the $\cal{I}$ components). The identities
in the strong component act like `strong charges'
and carry a relative mass sign. For example, in
the left hand matrix of (\ref{cq}) the two identities cancel
the two script identities leaving only the net 
5 q*'s to contribute to the strong component mass.

It is then a simple matter to 
calculate the basic current quark masses
in a proton;
these can be read straight  off expression ({\ref{cq}}).
There are 5 components in the proton
strong-interaction current quark part
and each couples to a C matrix of order 4!
The 4 identities in the E.M. component have
mass 4.4! and, after the cancellation
of a q and q* in the E.M. component
(these are considered to be opposite charges)
we are left with 3.(4!-2) as the order of the
3 q*'s in the E.M. component. There is a
doubling of order due to the fact
that the particle vector has two parity
permutations as indicated earlier 
in the text $(C,\bar{C})$ and $(\bar{C},C)$;

\begin{equation}
\mbox{Proton  current quark $T_r$ equivalent order}=
2\{9.4!+3.(4!-2)\}(1+\alpha_{em}+G_f)
\label{proton}
\end{equation}

\noindent
where  the same kind of radiative
gauge field correction as was done for
the other components is given.  This radiative
 correction  might
seem appropriate for the charged
components (the q's) but seems inappropriate
for the scalar identities which are
related to the strong interaction not the
electro-magnetic interaction. We will apply 
a $(1+\alpha_{em})$ correction to currrent quark mass
generated by both the identity (no em charge) parts
of (\ref{cq}) and (\ref{qc}) and the $q$ and $q^*$ parts
 just as  an em radiative
correction was applied to the constituent gluon energy in
(\ref{order3}) in relation to (\ref{glug}).
 Some justification for this is  found later in the text 
but it is quite speculative. The best that can otherwise be
said at this stage is that this approach seems to work in
all examples so far studied with this technique (which
includes a variety of mesons). Ultimately, however,
this situation is quite unsatisfactory and an understanding
of the radiative corrections to this theory of masses is
one of its' most important outstanding problems.

For the neutron the calculation is
analogous. The appropriate
matrices which define a colourless electrically-neutral
fermion are;

\begin{equation}
\mbox{strong component}=
\left(
\begin{array}{ccc}
\cal{I}&I&I\\
q^*&q*&I\\
q^*&I&q^*\\
\end{array}
\right),
\;\;\;
\mbox{E.M. component}=
\left(
\begin{array}{ccc}
{I}&I&I\\
q^*&q&I\\
q^*&I&q\\
\end{array}
\right)
\label{qc}
\end{equation}

\noindent
where it is implicit that the strong component
is coupling the the $C$ matrices of the particle vector
and that the E.M. component is coupling to the 
$\bar{C}$ components.
From these one can read-off the mass terms
applying exactly the same rules as before;

\begin{equation}
\mbox{Neutron current quark $T_r$ equivalent order}=
2.\{12.4!\}(1+\alpha_{em}+G_f)
\label{neutron}
\end{equation}

One can use eq's (\ref{proton}) and
(\ref{neutron}) to estimate the up and
down quark current masses but the
calculation is yet incomplete because 
there remains yet another component -
the $\omega$ component in the table at the
beginning of section  XII -
whose origin is something of a mystery.  This is calculated precisely
in the next section. When this is combined with
the above results very precise calculations
of the up and down current quark masses can
be made. It must be born in mind, however, that
these masses change somewhat according to context;
they are not exactly the same, for example,
in a pion as a nucleon. We will not calculate the
pion mass in this paper although the calculation follows
the same basic principles given here and values within
empirical limits are obtained; in fact the whole spectrum
of ground state spin-0 meson masses can be calculated 
but there is considerable complexity in calculations
involving 
higher-generation quarks which will not be discussed in 
this paper.

\section{glue energy associated with current quark masses}.

\begin{center}
\begin{tabular}{|c|c|c|c|}
\multicolumn{4}{c}{CURRENT QUARK U(1) COMPONENTS}\\\hline
quark type&`strong-charge'&generation multiplier&
strong component\\\hline
top&-&3&1.(4!-2)\\\hline
bottom&+&3&2.(4!-2)\\\hline
charm&-&2&1.(4!-2)\\\hline
strange&+&2&2.(4!-2)\\\hline
up&-&1&1.(4!-2)\\\hline
down&+&1&2.(4!-2)\\\hline
\end{tabular}
\end{center}

The above table lists the glue  corrections for
current quark masses for the three generations. (Why they
have been   
called U(1) components will be discussed later).
The glue  corrections for the up and down
quarks in the nucleons can be calculated from the strong
components in expressions (\ref{cq}) for the proton and
(\ref{qc}) for the neutron by `rotating' the strong component
to couple to the generators and setting
$\alpha_s=1$. (This `rotation' is an induced phase
change when they are moved from the left to right hand
matrix in (\ref{cq}) and (\ref{qc})).

A much easier way to consider  the current-quark glue
 corrections is with the leggo approach. Take two
Cantor cubes and stack them;

\vspace{0.5cm}

{\centerline
{\epsfig{file=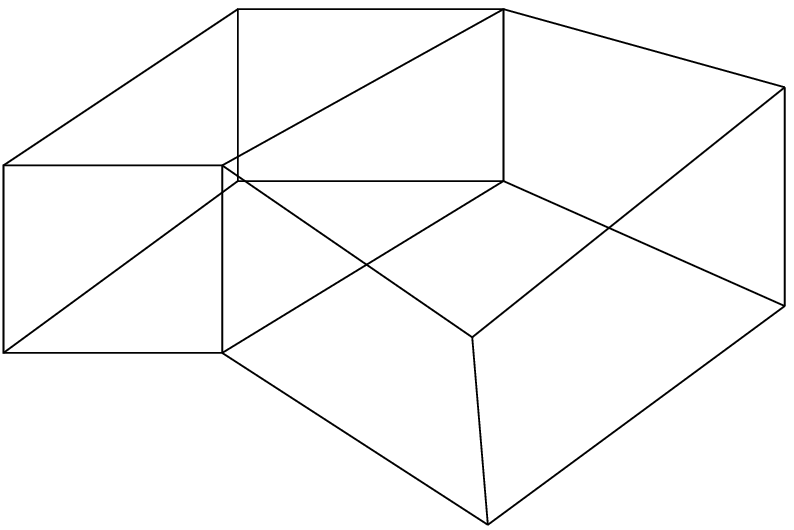,width=12cm}}}

\vspace{0.5cm}

\noindent
now, there are only 11 4-point sub-geometries in this
structure rather than the 12 in two separate cubes so one
4-point geometry has been lost in the stacking process.
The energy equivalence missing is equal to the group
order minus the generators;  here (4!-2) for
an object with spin. If we assume that the
cubes only stack on the non-triangulated parts of their
surface this means that the down-type quark has twice the
surface area for stacking compared to the up-type (this is
the source of the factor of two difference in the above table).
The strong coupling is strongly attractive  in the region $\alpha_s=1$
so energy has to be put into the system to pull them further apart
and there must be some potential energy of separation at this
energy scale. We will equate this with the `missing' energy in
the stacking model. The odd thing we have here, however, is that
the up and down quarks have a sign associated with the
non-triangulated parts of their topology; something quite 
unexpected for the strong force! This sign difference on the
identities is only found in the `strong' matrix in (\ref{cq}) and
(\ref{qc}) but is essential if the algebra is to be consistent.
Actually, the stacking energy is always positive but the 
up and down-types enter the calculation viz. the `strong' matrix
identities in (\ref{cq}) and (\ref{qc})
with a different sign. An example will illustrate
the ideas involved.

 The calculation for
the proton is found by summing over
all possible pairings of stacked
quarks using the
above table (gluons again expressed as $T_r$ equivalent groups);

\[
proton = |up+up| + 2|up+down|= |-2(4!-2)|+2|+(4!-2)|= 4(4!-2)\]

\[neutron= |down+down|+2|up+down|=|+4(4!-2)|+2|(4!-2)|=6(4!-2)
\]

\noindent
where the arbitrary  sign choice 
on the gluon components has been eliminated in the calculation.

Why glue  corrections do not originate from the
identities in the second block matrix (called the EM component)
in expressions (\ref{cq}) for the proton and
(\ref{qc}) for the neutron is unclear but comparison with 
empirical data shows that they do not contribute. It is possible
this is because the two block matrix representation of the
current quarks is fermionic and the single block rep. is appropriate
for the bosonic gluon  corrections.

\section{Nucleon masses continued}

It is reasonable to assert that 
quantum chromodynamics is well established empirically
as the correct theory  of the strong interactions. The 
coupling between the gluons and the quarks displays 
asymptotic freedom at high energy i.e. at high energy the
quarks behave like free partons and the coupling is weak.
At low energy however the 
coupling is strong and perturbative calculations are not
possible. Of necessity,  calculation of the rest mass
of any hadron will place the interaction in the non-perturbative
strong coupling region. 

In the previous sections a discrete model of QCD was
developed employing discrete rather than continuous colour
and calculations summing the matrix order 
of the discrete components of the model were made. The analogy 
with  the tetrahedral group is now to be extended to the
definition of mass. From a continuous group (SU(2) in the
case of the tetrahedron) a discrete sub-group of elements is
extracted which represents a finite {\it{bare}}
mass term where the  actual finite group 
involved is dictated by the symmetry of the geometry.
For affine-set geometries there is always a 
permutation symmetry of the points of the set.
 To calculate radiative corrections to masses, 
as in the case of the 
calculation of the rest mass of the electron (eq.{\ref{lepton}}),
we do not sum over boson momenta in the case of cubic
quarks. Instead
we invoke the same ansat{\"{z}} of equilibrating boson energy
to the Dirac mass of the associated fermion multiplied by 
the coupling constant at that energy scale. In the case at hand
one would expect that 
the scale appropriate is $q^2{\approx}m_q^{\;2}$
where $m_q$ is the current quark mass
but this may not be quite correct
because the current quark masses contain both
$q$/$q*$ charged pieces and $I$/${\cal{I}}$
electrically neutral pieces and the correct energy
scale to set $\alpha_{em}$ may be more like 
half the current quark mass. In any case
it makes little difference since $\alpha_{em\;q^2=m_q^2}$
is little different from $\alpha_{em\;q^2=0}$
for the up and down quarks.

 The value of the strong
coupling is specified by setting the gluon energy in matrix units
equal to constituent quark matrix units; 
this sets the strong coupling at 1 unambiguously  for the
non-perturbative calculation. 

We are now in a position to calculate our nucleon masses.
Basically we have four types of components. The first two
components are the constituent quark mass and the 
constituent gluon mass equivalence and these are
almost exactly the same (expressions 
{\ref{order}} and {\ref{glug}}). These terms, in the transition
to a field theory, represent the energy associated with
the motion of the current quarks and the momentum
of the gluons. They dominate the mass expression 
and indicate that, to a good approximation, about
1/2 the nucleon momentum is carried by the gluons. It
is possible to produce an exact calculation. The
constituent quark energy is;
\[6.(4!).6(4!-2).{\cal{R}}\]
and the constituent gluon energy is
\[6.(4!).6(4!).{\cal{R}}\]
where ${\cal{R}}=(\alpha_s+\alpha_{em}+G_f)$
and $\alpha_s=1$ precisely. 
 (A variable $\alpha_s$
coupling constant and asymptotic freedom will
only appear in the transition to a field theory
with all its attendant complications).

To these components we must add the current quark
masses which are in two pieces; the electro-magnetically
charged current quark masses calculated from the 
tables and the glue  corrections to the
current quark masses  calculated 
in the previous section. For the proton
we calculated the current quark mass
as;
\[2\{9.4!+3.(4!-2)\}.{\cal{R}}+4.(4!-2)\]
and for the neutron we calculated the
total current quark mass as;
\[2.12.4!.{\cal{R}}+6.(4!-2)\]
where in each case the glue   corrections
to the current quarks  are the
ones without electro-magnetic or weak radiative corrections. 
The derived current quark masses are up$\approx$4.6120MeV
and down$\approx$5.9053MeV.
One then obtains
the following expressions for the nucleon
masses directly;
\begin{equation}
\mbox{Proton Mass}\;={\cal{R}}.(8!-12)+4.(4!-2)
\;\;
;\;\;
\mbox{Neutron Mass}\;={\cal{R}}.8!+6.(4!-2)
\label{nuc}
\end{equation}

Notice that the cubic geometry is implicit 
in the resulting expressions even though each
quark is represented by an individual cubic geometry;
they sum to provide a single effective {\it{colourless}}
cube as is represented by the appearance of the
8! which is the order of the $S_8$ permutation
symmetry group. (The permutation symmetry is characteristic
of an affine-set geometry; in this case of 8 points).

 This seems to be  the reason
why the nucleons are so special;
they have the geometry of quarks but they are not
coloured. In this they appear to be unique
and it is possible this exceptional expression of 
symmetry is the reason why protons and neutrons are
so fundamental to physical structure. 

 The 12 massless generators 
in the proton mass expression are the
6 pairs of  generators associated with the surface of
the cube (they can be seen explicitly in the
current quark masses expressions; they are associated
with the three $q^*$'s in the E.M. matrix (right matrix)
of expression (\ref{cq}) - one $q$ cancels one of the
$q^*$'s and parity doubling doubles the number of 
$q^*$'s to 6 -  each of which is associated with two
massless generators of an $S_4$ symmetry;
the $q$'s in the right-hand matrix in (\ref{cq}) of course have
no generators)
and explicitly appear in the charged object
as would be expected.
The neutral cube, although its structure contains
analogous generators, does not manifest these
generators explicitly (the $q$'s cancel the $q^*$'s
in the current quark EM matrix (\ref{qc})).  The
mass expressions can be converted into
MeV by using the electron mass=${\cal{R}}(4!-2)$.
(This of course does not mean that the electron
has $\alpha_s$ in its gauge field correction;
the `1' in this case is the tetrahedral unit of
mass which is universal).

For simplicity we set $\alpha_{em;\;q^2=0}$
in the ${\cal{R}}$ for both nucleons and electron.
Setting $G_f\approx1.4\times10^{-5}$ 
(a value extracted from study of the lepton
masses - see later in the text) gives ${\cal{R}}\approx1.0073115$
and we obtain;

\[{M_p\over{M_e}}=1836.1528(1836.1526675(39)\]

and

\[{M_n\over{M_e}}=1838.6837( 1838.6836550(40)\]

Both values are a little high, but given the
uncertainties in the couplings $\alpha_{em}$
and $G_f$, the results are very
good.

\section{Why does the gluon momentum related to
the constituent quarks have a $\alpha_{em}+G_f$
radiative correction?}

Good question! As might be expected the gluon 
correction to the current quark mass only has a 
multiplier `1' which can be interpreted as $\alpha_s$,
but the `constituent' glue energy, (\ref{order3}) has an
electro-weak radiative correction! This appears quite
out of place; only an $\alpha_s=1$ should appear
with the glue momentum - but if we omit the electro-weak
radiative correction the result is well outside the
experimental value. The calculation results at the 
end of the last section clearly suggest that there
{\it{should}} be an electro-weak radiative correction
to the constituent glue energy, but why?

If we are to have a theory free of
any arbitrary parameters then the value of $\alpha_{em}$ must be
based on the geometry also. This is dangerous
territory (almost a taboo topic in physics)
but it must unfortunately be attacked.

There is only
one (Dirac) massive geometric object - the tetrahedron -
in two (fermionic) incarnations; as lepton and quark
(the cube geometry ultimately reduces to
tetrahedral units). The fine
structure constant appears to represents the ratio of the 
number of effective tetrahedral groups generated
by the  $T_c$ (cubic) generators compared with the single
tetrahedral geometry generated by the $T_r$
generators
(noting that in a unit charge cubic proton 
there is
 an amalgam of six effective pairs of  $T_c$
tetrahedral generators);
the lepton generators are about 137 times more potent
at generating electro-magnetic charge
than the quark generators. The explicit E.M. component
associated with the 6 pairs of $T_c$ generators on the surface 
of a unit charged cube such as a proton
 is 6.(4!-2) + radiative corrections. 
If we strip this off the proton and then remove the
constituent
gluon energy with only the $\alpha_s=1$ coupling to the
gluon energy - i.e. what we 
physically expect - (omitting the weak interaction in both cases),
 divide by six to get the relative charge-generating
power of one of the six pairs of $T_c$ generators and then
divide out the tetrahedral  units  we find
;
\begin{equation}
{1\over6}.{1\over4!}[
(1+\alpha_{em})\{8!-12-6.(4!-2)\}-\{6.4!.6.4!\}]
=
\alpha_{em}^{-1}
\label{alpha}
\end{equation}
\noindent
which gives a 
value  $\alpha^{-1}\approx137.03596$ 
which would be an appropriate value for $\alpha_{em}$
at the energy scale involved i.e. not at $q^2=0$ but
at the electron rest mass ($T_r$) scale of $\approx0.5MeV$.
Equation (\ref{alpha}) tells us that, as expected,
the gluons don't in fact have electro-magnetic radiative
corrections; the fact that the gluons carry no
electro-magnetic charge is absorbed into the value of
the coupling constant in the nucleon mass expressions
(\ref{nuc}).

Let is try and interpret this expression.
The proton as a cubic geometry has the
same geometry as a quark;
stripped of its current quark interactions and
gluon momentum  it
is  a large collection  of $S_4$ units with 4! matrix
elements each. 
 The surface charge 6.(4!-2)(1+$\alpha_{em}$) may be
considered an electro-magnetic interaction energy between the
charged current quarks.   Eq.(\ref{alpha}) tells us that,
when this latter energy, and the gluon energy, is eliminated,
one pair of $T_c$ generators generates exactly 137.0359...
 units of 4! matrix elements while the lepton pair $T_r$ generators
only generate one $S_4$ group of matrix elements
(we are not concerned with mass here but rather with
matrix elements; that at least is the justification
for using 4! instead of (4!-2) in the divisor
of eqn.(\ref{alpha}) - mass and charge of course  are not the
same thing!). So, in this sense,
the lepton $T_r$ generators are 137.0359.. times more powerful than
the quark generators and creating electro-magnetic charge.
Geometry generators generate intrinsic spin and so generate
the local (gauge) phase on the geometry  which generates the 
gauge fields which generate space-time.....

Thus maybe $\alpha_{em}$ 
 represents a compensating mechanism
which allows quarks and leptons to occupy the same space-time
with the same electro-magnetic gauge field coupling to
both; they have the same electro-magnetic vacuum.
All quite speculative to be sure but there must
be a reason for everything in nature yes? At least this
speculation can give some justification for 
the radiative correction found in eqn.(\ref{order3}).

 Once again the fact that the transition to a
formal field theory results in a variable $\alpha_{em}$
with energy scale does not invalidate this concept because
the masses also will change with energy in a field theory.
What we can assert is that the running coupling $\alpha_{em};(q^2)$
will evolve with energy in such a way the the quarks and leptons
continue to `see' the same space-time as the energy scale
evolves. A changing $\alpha_{em}$ indicates that the quarks
and lepton masses will not evolve in exactly the same way
with increasing energy.

Notice that there is an implicit assumption in 
eq.(\ref{alpha}) 
it
assumes that the electron and proton have exactly the
same absolute value of charge; i.e. charge quantisation.

\section{mass formulae with some general comments}

\begin{center}
\begin{tabular}{|c|c|}\hline
proton mass&{$\cal{R}$}.(8!-12)+4.$\omega$\\\hline
neutron mass&{$\cal{R}$}.(8!)+6.$\omega$\\\hline
electron mass&{$\cal{R}$}.(4!-2)\\\hline
muon mass&{$\cal{R}$}.(4!-2)+6.6!+2.5!\\\hline
tau mass&{$\cal{R}$}.(4!-2)+8!+7.7!+2.6!\\\hline
\end{tabular}
\end{center}

The value $\cal{R}$ here is considered to be
a non-perturbative radiative correction
which is assumed to have the following
simple form ${\cal{R}}=1+\alpha+G_f$
where $\alpha$ is the fine structure
constant and $G_f$ is the weak coupling constant
converted to a dimensionless constant by
use of a selected energy scale (we will
take a low energy scale whereby it is
assumed that, in comparison to the electro-magnetic
coupling constant, $G_f$ has a value $\approx10^{-5}$).
The extra bits added on to the muon and tau are
Higgs components and we will examine these more
closely later.

Of course $\alpha$ and $G_f$ are scale dependent parameters
whose value varies with energy scale but we are not
here necessarily implying  radiative corrections found by
summing over a range of boson energies. Rather it is
being assumed that, instead of a  perturbation expansion,
 a finite  radiative correction
can be applied to a finite `bare' mass; in the case of
electro-magnetism of order $\alpha$ at $q^2{\approx}m_e^{\;2}$
(although one must always keep an open
mind as it is possible the perturbation expansion
has a finite sum for some at present unknown
reason; that the 
electro-magnetic radiative correction to the electron
mass, 
for example,
is of order $\alpha$ has been suggested before{\cite
{sakurai}}.) 
The term $\omega$ has the value;
\[\omega=(4!-2)\]

There is one free parameter
in the mass expressions; the energy scale used to
render the weak coupling  to a dimensionless
number but the weak vector gauge boson
masses can be calculated. Notice that $\cal{R}$ is probably the simplest
possible conceivable 
non-perturbative radiative correction one could propose.

The best fit is given by  $G_f=1.4101\times10^{-5}$
using $\alpha^{-1}=137.035989$.
This then yields the following masses
(current empirical values in parentheses);

$M_{\mu}/M_e=206.7682621(206.7682657(63))$

$M_{\tau}/M_e=3477.4006381(3477.60(57)).$

Both ratios are within
one $\sigma$ of the empirical data and
are  predicted to higher
accuracy than current data and can thus be
tested. Since the predicted masses are all within
one $\sigma$ of the empirical value the $\chi^2$
value is less than one indicating a very good fit
with the data.

Can we discern any patterns in the other 
parts of the data formulas?
The obvious thing is the presence of factorial
numbers is all the expressions. This suggests a 
discrete symmetry under permutation. Evidently the
charged leptons are built around a symmetry 
that involves a permutation of four
objects (not necessarily particles!),
the nucleons a permutation of eight objects and
the pions seven with a specifically 
identifiable six-object subgroup
(the pions have not been included here). 

Suppose the basic lepton unit mass, ${\cal{R}}(4!-2)$, is
related to tetrahedral symmetry;
the permutation group of the vertices of 
a tetrahedron is isomorphic to the discrete
group $T_d$ of the three-dimensional tetrahedron.
This  group has two generators and 4! elements.
An alternative more realistic choice may be $T_r$
the unique 24 element sub group of SU(2)
which also has two generators;
the -2 can then be considered to correspond to the
generators of the group (the masslessness
of which we associated with
intrinsic spin).  This sub-group is of the
non-abelian type. The tetrahedral structure may
represent a type of 3-dimensional brane
or orbifold. 
The idea of such a symmetry
underlying physical structure is of some current
interest in field theory and of course has
long been a feature of physical chemistry
(tetrahedral symmetry is a feature of some crystal
structures for example). Firstly it has been pointed out that
certain soliton models of the Skyrme type
(albeit for baryon structure) have symmetry
patterns of a tetrahedral, cubic and icosahedral
nature \cite{Sut}. Secondly there has been some
discussion of tetrahedral symmetry in relation
to orbifold gauge field theory 
(for example \cite{Douglas}, \cite{Johnson},
\cite{Hanany},
\cite{Greene} and \cite{Muto}; for a discussion of
permutation symmetry in relation to orbifolds
see \cite{Klemm} and \cite{Bantay}).
Of interest for example might be the 
association of tetrahedral symmetry with a 
discrete subgroup of SU(3) which encompasses
gauge fields appropriate for the standard
model \cite{Hanany} ${\hat{E}}_6\approx{T}$
and ${\hat{E}}_7\approx{O}$ where O is the octahedral
(cubic) discrete symmetry.
We could then try to associate the 8! elements
in the nucleons with a cubic type of symmetry.
This interpretation works rather neatly for 
the neutron because the extra $\omega$ term
in the mass expression for the neutron -
the 6(4!-2) - could be interpreted as related
to the six square faces of a cube each of
which could be allocated a tetrahedral-equivalent
symmetry of total 4! elements with two generators
(each square face of course has four vertices).
The appearance of electro-magnetic radiative 
corrections for the neutron might also be taken
to imply charged sub-structure.

Of course what is being studied in the above 
references such as \cite{Douglas}, \cite{Johnson}
and \cite{Hanany} is, chiefly, the issue of symmetry
(and in particular supersymmetry)
groups connecting different {\it{types}}
of particles not masses of individual particles.
However, there may be a connection. We might, for
example, view the discrete symmetry as a kind
of symmetry of the unobservable {\it{bare}}
particle; a particle devoid of its investing
gauge fields. Of course the `bare' particle
structure in field theory is not only
unobservable it is mathematically
inaccessible being plagued by infinities
(infinite mass, infinite charge etc.)
but this might be an artifact of our ignorance
of some underlying discrete structure. The surface geometries of 
such a discrete structure   may form the gauge fields
as a worldvolume brane type structure
rendering the underlying symmetry unobservable.
For example, if the isolated tetrahedron 
is a `bare' lepton then the triangular surface
geometries may be considered to be photons
and the individual interval `edges' of the geometry
to be gravitons. (A quick review of the
geometry of these objects will reveal that
the triangle symmetry is appropriate for
a discrete spin-1 object and the single interval
or `edge' of the geometry appropriate for a
spin-2 object).  Because of the non-abelian
nature of the double-cover tetrahedral group
its overall discrete `SU(2)-phase'
might then be `gauged-away` by the weak vector
gauge bosons so that all phase information 
in relation to the discrete geometry -
the phase of each individual edge, the U(1)
phase of each investing triangle and the
global SU(2) phase of the total
geometry, becomes experimentally unobservable
when the gauge fields propagate and 
sweep-out world-volumes that render the 
underlying discrete symmetry unobservable.
Such a scenario intrinsically demands 
{\it{local}} gauge invariance because of the
discrete nature of the geometry (perhaps
this is why nature chooses local gauge
invariance?).
A flat `square' version of the tetrahedron
would be necessary to account for uncharged
neutrinos of course; the absence of any triangular
sub-geometry would then correspond to the absence
of electric charge - it would of course retain
an overall SU(2) weak `local' phase and its
gravitational investing gauge field associated 
with the individual investing intervals of the
geometry. There are problems with this interpretation
however, because a flat two-dimensional geometry
is nominally spin 1 and massless. We will return to
this issue later.

The only residue  of discreteness would then be the rest
mass of the object.
The symmetries between particle types may then
arise because of this shared underlying discrete
structure; we can see from the mass formulas
for the hadrons that the tetrahedral symmetry
appears to be embedded in the cubic symmetry for example.
There is in fact quark-lepton unification here.

\section{vector gauge boson masses and electro-weak mixing}

Now, $T_r$ (or the 3-geometry 
analogue $T_d$), is  spinorial  in the double
covering sense so that this group
would be appropriate for a spinor particle but not
for a boson. The analogous `bosonic'
group for a tetrahedron is T which has 
12 elements and two generators. Suppose we 
were to choose this kind of geometry for 
our vector gauge bosons, since it matches
 the geometry of the leptons 
in this model apparently and is triangulated;
a requirement for charged vector gauge bosons
at least. How can we
create a factorial expression out of the
T elements; assuming that this is the only
way mass is expressible for an affine-set
geometry? The easiest
and simplest 
way is to take the non-generator elements
of T 
and assemble them into a permutation 
group viz;
\[
(T-2)!=10!
\]
and we can then arrange, viz some
appropriate choice of decomposition,
a model for the vector boson masses.
First, note that any permutation number 
has a unique decomposition into a sum of
permutations;
\[
n!=(n-1)(n-1)!+(n-2)(n-2)!+...+1.1!+0!
\]
so if we take the 10! elements and strip-off
a tetrahedral equivalent unit 4!=3.3!+2.2!+1!+0!
to couple to the underlying four-point  geometry (and define
it physically - a reasonable pre-requisite) we end up with
a unique and inevitable decomposition into  six components
if we are working with affine-set geometries;

\[10!=9.9!+8.8!+7.7!+6.6!+5.5!+4.4! +(4!)\]

\noindent
 and we throw away the last 4!
Each `10!' should correspond to the mass of one
vector gauge boson. At this stage comparison with the
empirical data shows that
 maximal mixing of 
the {\it{charged}} components on the 8! (cubic)
and 4! (tetrahedral) 
geometry terms occurs. The charge appears only in the $W^{\pm}$
so the mass shifts only to the $Z^0$ in maximal mixing;

\begin{center}
\begin{tabular}{|c|c|c|c|c|c|c|}\hline
Boson&9.9!&8.8!&7.7!&6.6!&5.5!&4.4!\\\hline
$W^+$&1&1/2&1&1&1&1/2\\\hline
$W^-$&1&1/2&1&1&1&1/2\\\hline
$Z^0$&1&2&1&1&1&2\\\hline
\end{tabular}
\end{center}

\noindent
which sums to 3.10! as required.

 In these mass expressions
what has happened is that each
$W^{\pm}$ has `donated' 
half its' charged $S_8$ component
of mass to the $Z^0$; a $4.8!(1+\alpha_{q^2=M_W^2})$
from its mass to the $Z^0$ so that the $Z^0$
has a term $(8.8!+ (1+\alpha_{q^2=M_W^2})8.8!)$
instead of just 8.8!  as its
second component of mass; and similarly for the
$S_4$ terms. 
In this process we expect each `donated' piece to  keeps its'
radiative correction. 
On the basis of the analysis of 
$\alpha_{em}$ in section Taking $\alpha=128^{-1}$
`on-shell'
at $q^2=m_W^2$ instead of $q^2\approx0$
we obtain;

\[M_z^0=91.1729
(91.188(0.007))GeV\;\;\;\mbox{and}\;\;\; M_W=80.5794(80.49(0.14))GeV\]
\noindent
which gives, using $cos\theta={{M_W}\over{M_Z}}$,
$sin^2\theta=0.2188$. This is the value with
radiative corrections to the masses included.
If one omits these one easily obtains $sin^2\theta=0.2299$.
It is interesting to note that the weak mixing
is related to shifting of mass components from 
both the charged bosons to the neutral vector
gauge boson. For example, if this mixing did
not occur the weak mixing angle would be very small
and due only to radiative corrections to the 
W mass.  Notice that charge
neutrality of the $Z^0$ is maintained in spite
of the radiative corrections to its mass since
it obtains two equal contributions from opposite
charge consignments.

What is most interesting is that 
1. the electro-weak mixing is taking 
place at the $S_8$ symmetry level;
exactly the level we used previously to
extract a value for the fine structure
constant (the proton mass is an $S_8$
symmetry)and this is the reason for adding
the $S_4$ mixing even though on the basis
of the empirical data this is not so strongly
implied, and
2. the mixing is maximal! There is absolutely
no arbitrariness at all in the value of the
weak mixing angle if one can account for the
maximal mixing (which unfortunately we cannot).

\section{the Higgs field}

Let us focus first on the masses of the
charged leptons. Notice that the component
with radiative corrections is identical
in each of the three species of lepton.
This suggests that the component carrying the
electro-magnetic charge is identical in each
case and this is consistent with charge
quantisation. 
The formulas for the leptons imply that there
is a common component in the leptons which
is responsible for generating the electro-magnetic
charge (and weak interaction properties) of
the individual leptons as well as the 
intrinsic spin (it is the only piece
with generators set explicitly massless). Let us then
attribute the remaining mass components 
in the muon and tau to the Higgs scalar field.
This is a spinless field and the lack of
a radiative correction for these mass components
(at least to the level of the current data)
implies that, if these components are the result of
the Higgs mechanism alone, that the Higgs particle is 
not electro-magnetically charged because of the
absence of any  associated radiative correction 
for the electro-magnetic field. 
 We might also postulate
that the -2 appearing with the charged component 
of lepton mass is
related to intrinsic spin. This makes some sense
since, firstly, it implies identical intrinsic
spin in all three massive leptons and, secondly, its absence
in the components in the lepton masses without
radiative corrections would then imply that they are
spinless as required of the Higgs field.
Whether the Higgs can also be
considered responsible for the ${\cal{R}}(4!-2)$
part of the lepton mass is an open question.
 This piece of mass, identical to the electron rest mass,
 we might attribute it to the 4! piece we `throw' away 
when we decompose a large scalar factorial unit. Clearly we need
the Higgs for correct dimensions in the Lagrangian;

\[
{\cal{L}}=g\psi_R\psi_L\]

\noindent of any fermion.

This prompts a search for the pattern of Higgs components
appropriate for the  massive quarks and leptons. For consistency
we will use the same basic structure that we employed for
generating the Higgs components for the vector gauge
bosons since the same Higgs generates the masses for
both sets of particles.
 From the known spectrum of masses clearly the
symmetry will be broken. Using the components already
listed for the tau and the muon the symmetry breaks up
as follows (using a $2.S_{10}$ to couple to the fermions
in comparison to the $3.S_{10}$ we used to couple to
the adjoint representation of the vector gauge bosons);

\begin{center}
\begin{tabular}{|c|c|c|c|c|c|c|}\hline
Fermion&9.9!&8.8!&7.7!&6.6!&5.5!&4.4!\\\hline
$top$&2&10/8&0&0&0&0\\\hline
$bottom$&0&5/8&0&0&0&0\\\hline
$charm$&0&0&1&3/6&0&0\\\hline
$strange$&0&0&0&1/6&3/5&0\\\hline
$up$&0&0&0&0&0&0\\\hline
$down$&0&0&0&0&0&0\\\hline
$\nu_{\tau}$&0&0&0&0&0&0\\\hline
$\tau$&0&1/8&1&2/6&0&0\\\hline
$\nu_{\mu}$&0&0&0&0&0&0\\\hline
$\mu$&0&0&0&1&2/5&0\\\hline
$\nu_{e}$&0&0&0&0&0&0\\\hline
$e$&0&0&0&0&0&0\\\hline
\end{tabular}
\end{center}
\noindent

The known quark masses have been used as a guide
to the decomposition. The first generation quarks and leptons
have been left massless (and consequently 
an amount of mass equal to 5.5!+8.4! unaccounted for!).
The pion mass expressions however imply that, under some
circumstances, the up acquires a contribution of 5!
and the down of 4!. Detailed study of meson ground state
masses (unpublished\cite{filewood}) appears to indicate
that these missing components - the 5.5!+8.4! - do
not appear in any of the second or third generation quarks.
It is possible they contribute to the $\omega$ terms that
appear in the hadron mass expressions or more likely they
in fact manifest as the baseline tetrahedral units in
both quarks and leptons (for example the (4!-2) part of
the electron mass). As mentioned, such a contribution
from the Higgs is needed to get the dimensions of the
Lagrangian right for all Dirac particles.
From the table the Higgs contribution to the quark
masses are; top $\approx159.9$GeV, bottom $\approx4.65$GeV,
charm $\approx863$MeV and strange $\approx24.9$MeV.
It is interesting to note that the fifth flavour
line appears to be breaking up in SU(5) multiplets and the
third quark flavour line in SU(3) multiplets. The
lepton terms are dominated by singlets.

With these postulates in mind can we make a 
guess at the Higgs mass?
Obviously it must be related to a factorial
number if these mass expressions have any
real physical meaning.
The minimum choice would, from the above
discussion, be a multiple of 10! which
translates into an integral  multiple of about 83GeV.
However this is based on the (T-2)! bosonic geometry
not on a scalar geometry. Could we propose a
scalar geometry such as T! (which would
be scalar under the assumption of no massless
generators)? Unfortunately this
translates into a mass of the order of 11.05TeV
which appears to be an order of magnitude too
high for a prospective Higgs scalar which
is expected to be of the order of 1TeV or less.

\section{neutrinos}

The theory of neutrinos has been left
to last in this paper because this is 
probably the most difficult area of the
theory to analyse.

Let us first consider what the geometry of
a neutrino {\it{should}} be from first 
principles and see where the problem lies.
The massive leptons in this theory are
particles which define a tetrahedral
3-space continuum. The different masses
of the e, $\mu$ and $\tau$ derive solely
from the different Higgs components
listed in the previous table (Section XV). 

The Higgs components are considered to be
`internalised' in the tetrahedral geometry;
that is, they are located {\it{inside}}
the boundary of the geometry in a time-less
space. The number of actual vertices in the
geometry depends on the leading term of 
the Higgs mass components. Thus, the tau
lepton, which has components 8! + 7.7! +2.6!
is an 8-vertex geometry with 4 vertices internalised
inside the tetrahedron (the 8! mass component relates
to a $S_8$ permutation symmetry of the vertices).
The boundary of the geometry which embeds in the
surrounding space-time is identical to the electron
and has a tetrahedral morphology.
The $S_6$ and $S_7$ symmetries are sub-groups of
the 8-vertex geometry. Similarly we can see,
from the Higgs components, that the muon must be
a 6-vertex geometry with two internalised vertices.

Let us hypothesise that the corresponding neutrinos
conserve vertex number; this is geometrically the
most intuitively obvious choice. This means that
the $\nu_{\tau}$ is an 8 vertex 
(but two-dimensional) geometry, the 
$\nu_{\mu}$ is a 6-vertex (two-dimensional)
geometry and the $\nu_{e}$
a 4-vertex geometry flat `square'.
The proposed geometric forms are all two-dimensional
and are as follows;

\vspace{0.5cm}

\epsfig{file=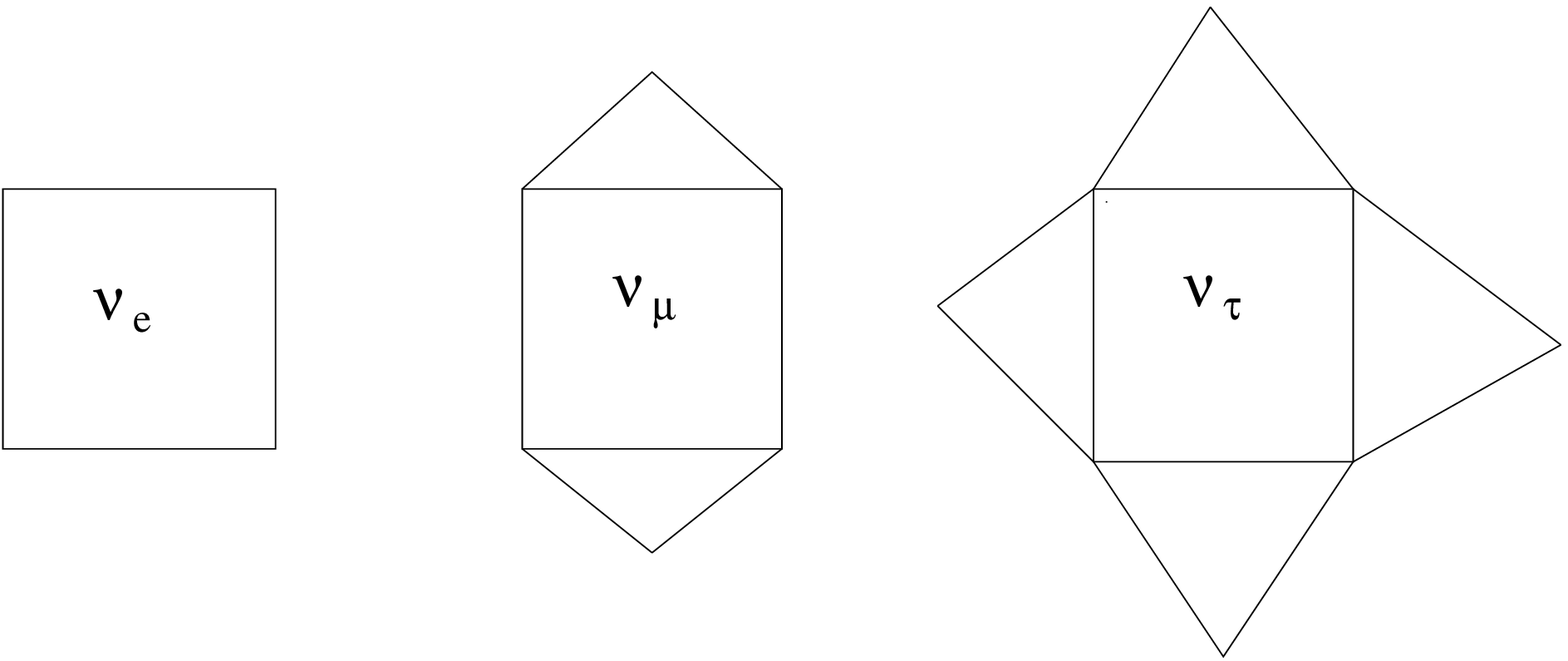,width=15cm}

\vspace{0.5cm}

 The absence of any Higgs components
for mass generation in the neutrinos means that they
cannot be Dirac particles in this theory. 
Let us see why.

It is instructive to consider first the case of 
the electron neutrino.
The electron has a basic mass unit formed from
the tetrahedral symmetry group because it forms
a 3-volume in space-time. This mass is given
 by a
tetrahedral  $S_4$
symmetry. The corresponding neutrino must be
a 4-vertex geometry but morphologically distinct
to the electron. Because in affine-set geometry
all tetrahedrons are equal, the unique choice would appear to
be a flat 4-vertex square. But this {\it{must}}
be massless in this theory as has been previously
explained. Worse still, such a geometry represents
a boson! (Unlike with the case of the tetrahedron,
no spinor group can be associated with a two-dimensional
square symmetry as the geometry generator acts as a vector). 

There appears to be only one way around this 
impasse and that is to assume that the 
structure of space-time is more complex than
has been listed in the hypotheses so far. 
We add the following;

Hypothesis 11;
That the time dimension for spinors is
retarded with respect to bosons.

This is analogous to admitting a violation
of special relativity into the theory but
with special restrictions. What it means is 
that the neutrino can be a flat square geometry
with respect to time as defined for an electron
but remains a tetrahedral geometry with respect
to a photon and time as defined for a macroscopic
observer; so it will be a spinor with
respect to any boson interacting with
it. For this to be achieved we need two
copies of the Lorentz group; one for spinors and
one for bosons (surprisingly CPT symmetry between
particles and anti-particles will be maintained
in such a scheme because the symmetry is broken
between bosons and fermions; not between two
fermions so that particles and anti-particles
would still have the same mass and lifetime etc).  

In terms of cardinality this means admitting the
existence of a cardinality $\aleph_1$ such that
\[\aleph_0<\aleph_1<c\]
as permitted by the formal undecidability of 
the continuum hypothesis. This means that, in the 
absence of the weak interaction, the continuum hypothesis
applies to the structure of the universe but the
addition of the weak interaction, and specifically 
the Higgs field, is equivalent to adding an
extra order of cardinality to space-time.
The propagating neutrino sweeps out a space
of cardinality $\aleph_1$ which is less than
that of the continuum. Its' weak radiative corrections
must then be responsible for the elevation of the
cardinality of the contained space to that of the
continuum and in so doing generate a
mass for the neutrino. A good geometric way to think
of this is as follows. In this theory gravitation
is the gauge field of individual intervals in the
boundaries of geometries - one dimensional `phase'-
and electro-magnetism is
the gauge field of triangles in the boundaries of
geometries - a two-dimensional gauge field (a U(1) phase in
this case).
 The weak interaction is like a `volume' gauge
field; a three-dimensional phase of the 
geometry and, because this dimensionality is
associated with mass in this theory, the corresponding
bosons must be massive.
 
In the case of the electron,
the contained 3-volume within the tetrahedral boundary
in the absence of the weak interaction is a space
of cardinality $\aleph_1$ which weak radiative
corrections elevate to cardinality c. The weak interaction
pushes up the cardinality of the contained space within
the tetrahedron.
The mass gap generated for the electron is
simply the weak radiative correction to its'
mass which was previously calculated at about
7eV. We expect a similar mass gap between the 
electron neutrino and the photon but this value 
seems to be ruled out by experiment. Nevertheless,
in the absence of any other physics modifying things,
this value is a definite prediction for the 
$\nu_e$ mass in this theory; a little under 
7eV ($\approx6.95$eV).

In fact, looking at the lepton mass expressions
given in Section XV  it is easily seen that the masses of the
corresponding neutrinos are degenerate all
with mass 6.95eV in the absence of any 
physical properties associated with the
`triangle' `wings' on the $\nu_{\mu}$
and $\nu_{\tau}$ in the previous figure
(these are expected to be scalar, not
spin-1, and so should not imply fractional
EM charge - the only other possibility is
that the triangles are paired with opposite
spins in which case both the $\nu_{\mu}$
and the $\nu_{\tau}$, but not the $\nu_{e}$,
 will have non-zero magnetic
moments).

A possible source of  non-degeneracy of mass
of the neutrinos  may be 
derived from radiative corrections to the
Higgs components in this theory which 
have been assumed to be zero
(the only multiplier to the factors
6.6!+2.5! - the assumed `Higgs' components
of the muon for example - is the number 1;
no electro-magnetic or weak radiative correction
appears). Using the available
data it is possible to put theoretical constraints
on  mass differences
arising from radiative corrections to the
Higgs field (which would inply non-standard
physics). One readily obtains
from the above table and the standard deviations
of the empirical values;

\[
Mass\;\;\nu_{\tau}-Mass\;\;\nu_{\mu}<0.7MeV
\]

\[
Mass\;\;\nu_{\mu}-Mass\;\;\nu_{e}<9eV
\]

to 95{\%} CL. As the charged lepton masses
become known more precisely these constraints can be
further restricted.

Can we have sterile neutrinos?

If the triangular `wings' on the $\nu_{\mu}$
and $\nu_{\tau}$ are spin 0, and if there
is a mass difference between states to provide
some phase space, then it should be possible
to produce the following weird geometries
through decay mediated by a virtual  Higgs particle;

\vspace{1cm}

{\centerline
{\epsfig{file=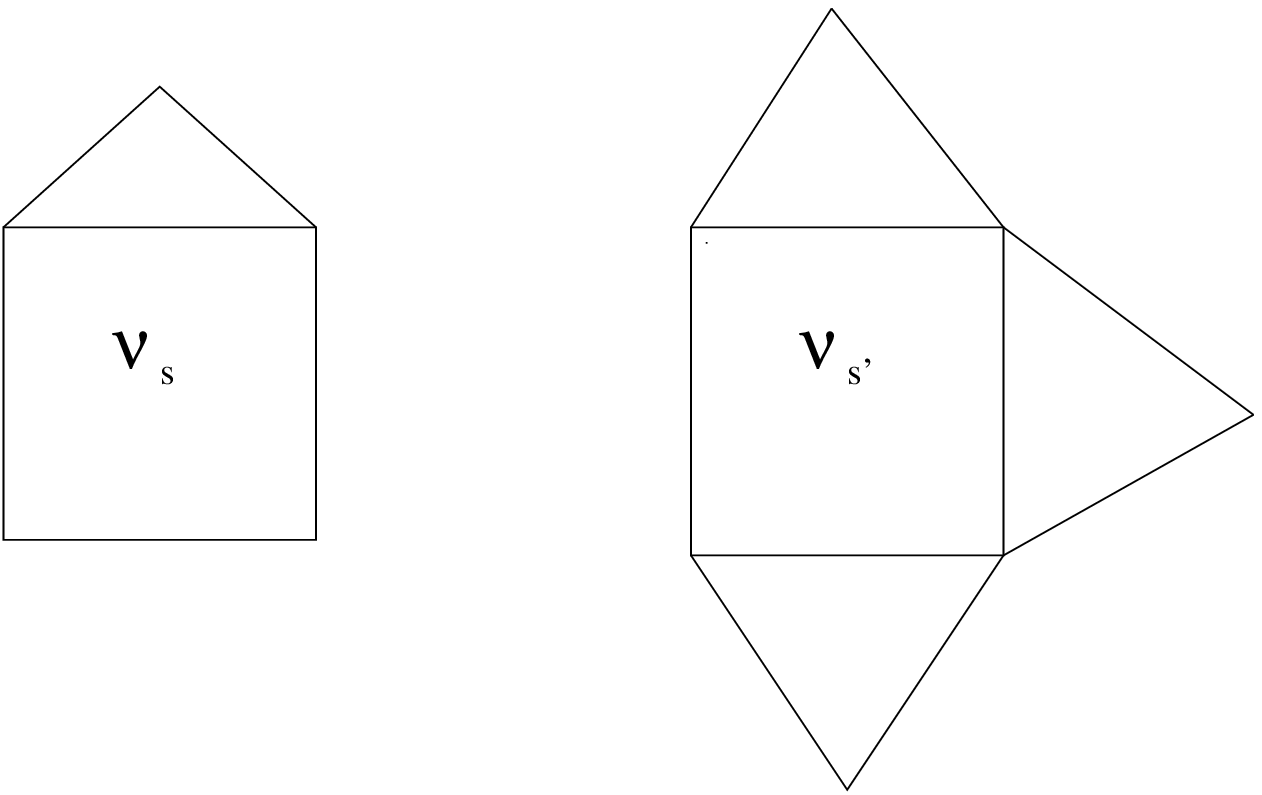,width=8cm}}}

\vspace{1cm}

\noindent
both of which have zero lepton number and in turn
would decay to the electron neutrino. If the
triangular wings are pairs of spin-1 these states 
cannot exist but instead a  variety of
anomalous non-standard interactions between neutrinos and
matter must be  possible causing interconversions of
one neutrino form to another; always complicated by the fact that,
in this model, special relativity will also break down
in the neutrino sector.

\section{conclusion} 

In summary mass expressions for all the quarks,
all the leptons and the gauge bosons (under the
assumption of massless gluons and photons) have
been presented which are concise and simple
and all related to a permutation symmetry
embedded in affine-set geometry.
In particular, precise mass expressions have 
been presented for the leptons, the nucleons
 and the vector gauge
bosons all of which are testable in precision 
experiments. Theoretical restrictions have 
been placed on the possible values the Higgs 
particle mass can be in this schema; if
it exists at all as a free particle 
(and in this theory there is reason to believe that
it may not!) it should
appear as an integer multiple of about 83.67GeV.
The most likely candidate is a multiplier of 
5 (to account for the `3' of the vector gauge
bosons and the `2' of the fermions). However, the 
most logical value for the Higgs is in fact
the true tetrahedral scalar boson  12! but this is
more than 10 TeV and so  appears to be ruled out.

The best precision test of the theory is
the tau lepton mass which, in the absence of
radiative corrections to the Higgs field,
can be calculated precisely using the known
values of $M_e$ and $M_{\mu}$. The current
prediction is $M_{\tau}=1776.95$MeV.

\end{document}